\documentclass[11pt,letterpaper]{article} 
\usepackage[includeheadfoot,
            marginratio={1:1,2:3}, 
            width=412pt, 
            height=688pt,]{geometry}

\usepackage{marginnote}

\usepackage{amsmath}
\usepackage{mathtools}
\usepackage[all]{xy}
\newcommand{\be}{\begin{equation}}
\newcommand{\ee}{\end{equation}}
\newcommand{\bea}{\begin{eqnarray}}
\newcommand{\eea}{\end{eqnarray}}
\newcommand{\mbb}{\mathbb}

\newcommand{\mc}{\mathcal}

\newcommand{\K}{\mc{K}}

\usepackage{jheppub} 

\usepackage[T1]{fontenc} 
\usepackage[utf8]{inputenc}
\usepackage{placeins}

\usepackage[english]{babel}

\usepackage{graphicx}
\usepackage{dcolumn}
\usepackage{bm}
\usepackage{placeins} 
\usepackage[english]{babel}
\usepackage{amsmath,amssymb,amsfonts, bm,bbm,slashed, subdepth}
\usepackage{graphicx}
\usepackage[sort&compress]{natbib}
\usepackage{xcolor}
\usepackage[normalem]{ulem}
\usepackage{hyperref}
\usepackage{cleveref}
\definecolor{red}{rgb}{1.0, 0, 0}
\usepackage{enumerate}
\usepackage{epsfig, subfigure}
\usepackage{setspace}
\usepackage{booktabs, tabularx}
\usepackage{units}
\usepackage{placeins}
\usepackage{multirow}
\usepackage{mathtools}

\usepackage[colorinlistoftodos,prependcaption,textsize=tiny,obeyFinal]{todonotes}

\usepackage{amsmath,blkarray,booktabs,bigstrut}

\newcommand{\nc}{\newcommand}
\nc{\lb}{\llbracket}
\nc{\rb}{\rrbracket}
\nc{\gl}{\llbracket}
\nc{\gr}{\rrbracket}

\newcommand{\eq}[1]{\begin{equation}
                     \begin{split} #1 \end{split}
                     \end{equation}}

\newcommand{\lab}{\mathsf }
\newcommand{\op}{\hspace{1pt}}

\allowdisplaybreaks[2]
\numberwithin{equation}{section}

\title{Large Volume Scenario from Schoen Manifold with de Sitter under Swampland Conjecture}

\author[a,1]{Rui Sun}

\affiliation[a]{Korea Institute for Advanced Study,\\
85 Hoegiro, Dongdaemun-Gu, Seoul 02455, Korea}

\emailAdd{sunrui@kias.re.kr}

\abstract{
To naturally allow for string compactification with duality manifested, here we investigate in the self-mirror large volume scenarios from Schoen Calabi-Yau manifold. 
We explicitly study the geometry of Schoen Calabi-Yau threefold and complete its triple intersection from both ambient and non-ambient spaces. Based on these, we study the large volume scenario of self-mirror Calabi-Yau compactification with Schoen type. 
Moreover, by studying the leading non-perturbative terms of the effective scalar potential, we find special uplift terms in order of F-term $\mathcal{O}(\frac{1}{\mathcal{V}^2})$ arising from self-mirror large volume scenario.
In particular, the quotient Schoen and Schoen Calabi-Yau large volume scenarios both give rise to de Sitter vacua.
In addition, we discussed on the criteria to the effective scalar potential derived from self-mirror large volume scenario according to the swampland conjecture with the constraints fulfilled. 
}

\begin{document}

\maketitle

\section{Introduction}

String compactification to de Sitter vacuum has been a challenging problem with the prerequisites from swampland conjectures  \cite{Ooguri:2006in, Garg:2018reu, Ooguri:2018wrx, Palti:2019pca, Lust:2019zwm, Bedroya:2019snp}. Although with duality manifested, the developments of double field theory and non-geometric flux compactifications provide one approach to derive de Sitter vacua, the large volume scenario for the non-geometric framework became a new issue. 
Since the non-geometric fluxes are supposed to arise from the T-dual framework of the standard large volume scenario, whether the high order $\alpha'$-correction and non-perturbative correction can still be suppressed at large volume limit become an important question. 
On this perspective, the large volume scenario with duality manifested where we are allowed to safely suppress the high order $\alpha'$ and non-perturbative corrections remain to be investigated. 
As one solution to this issue,  
here we turn to study the self-mirror large volume scenario from Schoen Calabi-Yau manifold whose mirror dual is itself and naturally with duality manifested.

In the general case, it has been shown in \cite{Balasubramanian:2005zx, Conlon:2005ki} that higher order $\alpha'$ corrections and non-perturbative instanton corrections are suppressed in Type IIB flux compactifications with Calabi-Yau hodge numbers $h^{1,2}>h^{1,1}>1$. Many interesting Calabi-Yau manifolds with relatively small number of K\"ahler moduli, such as Swiss cheese and Quintic type were studied in Calabi-Yau compactifications.  
The non-geometric phases usually lead to a small volume regime where $\alpha^\prime$ tree-level correction became important as investigated for quintic Calabi-Yau manifolds in \cite{Blumenhagen:2018nts} which approaches the stringy regime in Landau-Ginzburg phase. Together with the large volume limit, a hybrid regimes were studied for refined swampland distance conjecture.
Along the limit of finite volume and non-geometric background study, in \cite{Lee:2019wij}, Lee, Lerche and Weigand classified the possible infinite distance limits in the classical K\"ahler moduli space of a Calabi-Yau threefold while each limit at finite volume is characterized by a universal fibration structure. And the generic fiber shrinking in the limit is either an elliptic curve, a K3 surface, or an Abelian surface. Non-geometric Calabi-Yau backgrounds and K3 automorphisms were also discussed  in~\cite{Hull:2017llx}. 

On the other perspective, the Calabi-Yau compactification with both geometric and non-geometric fluxes tend to involve both the large volume limit and large complex structure limit at the meanwhile. 
To be in the large volume and large complex structure limit in one specific Calabi-Yau compactification, we are lead to study the special self-mirror Calabi-Yau compactification. 
Self-mirror Calabi-Yau manifolds are constructed with whose mirror duals are themselves. Such as in \cite{Braun:2011hd}, when one construct quotient Calabi-Yau from 24-cell self-dual polytope, generic Calabi-Yau hypersurface in the toric variety can be obtained.  The Hodge number of a smooth quotient $X=\tilde X/G$ Calabi-Yau threefold can be studied via  $G$-invariant polynomials as well. 
As one special self-mirror Calabi-Yau, Schoen Calabi-Yau manifold played a particular interesting role in comparing the $\mc{N} = 1$ F-theory compactification over $d P \times P^1$ from Donagi, Grassi and Witten's work to  their dual heterotic models via modular superpotential \cite{Curio:1997rn, Curio:1998bv}.
In which, the Schoen threefold with $h^{1,2}=h^{1,1}=19$ is constructed with fiber products of rational elliptic surfaces~\cite{Schoen:1988, Braun:2007sn}.
Its ${\mathbb{Z}}_3 \times {\mathbb{Z}}_3$ quotient with fundamental group $\pi_1(X)=\mathbb{Z}_3\times \mathbb{Z}_3$ were utilized to construct heterotic standard model such as in~\cite{Braun:2004xv, Braun:2005zv, Braun:2005ux, Braun:2005nv, Braun:2007sn, Donagi:2000zf, Ovrut:2018qog}.
Other interesting self-mirror Calabi-Yau manifolds were also constructed from different fundamental groups with smaller hodge numbers, such as in \cite{Candelas:2008wb, Braun:2007sn, Braun:2009qy}.

For Schoen Calabi-Yau manifold, flux compactification has been challenging due to the non-ambient divisor/space involved. 
Alternatively, quotient Schoen Calabi-Yau compactification which only involves the ambient divisors were utilized in heterotic M-theory compactification \cite{Deffayet:2023bpo, Deffayet:2024hug} and in self-mirror large volume scenario study \cite{Sun:2024ral}.
In this work, we firstly study the detailed geometric construction of Schoen Calabi-Yau threefold with explicit triple intersection derived from both ambient and non-ambient spaces. 
As follows, we study the self-mirror large volume scenario from both quotient Schoen and Schoen Calabi-Yau threefolds with de Sitter vacua obtained from order $\mathcal{O} (\frac{1}{\mc{V}^{2}})$ uplift term, which is in the same order as the  F-term $\frac{D W. DW}{\mc{V}^2}$. In addition, we discussed on the related phenomenology, cosmology and swampland conjecture with the constraints satisfied.

\section{Calabi-Yau Flux Compactifications}

In Calabi-Yau flux compactification, 
we first review the basics of Calabi-Yau geometry and then present the compactification with fluxes. For more details, we refer to \cite{Grimm:2004uq, Benmachiche:2006df, Grana:2006hr, Blumenhagen:2015lta}, and references therein.
In general, the symplectic basis for the middle cohomology of a Calabi-Yau threefold $X$ can be denoted by
\eq{
  \label{basis_001}
  \{\alpha_{\Lambda},\beta^{\Lambda}\} \in H^3(X) \,, \hspace{30pt} \text{with}~
  \Lambda =0,\ldots, h^{2,1} \,,
}
while the only non-trival pairings of the basis read
\eq{
  \label{symp_01}
  \int_{X} \alpha_{\Lambda}\wedge \beta^{\Sigma} = \delta_{\Lambda}{}^{\Sigma} \,.
}
In the basis \eqref{basis_001}, one can expand the holomorphic three-form $\Omega$ by
\eq{
\label{hol_three}
\Omega=X^\Lambda\, \alpha_\Lambda - F_\Lambda\, \beta^\Lambda,
} 
where $X^\Lambda$ and $F_\Lambda$ are the periods as functions of the complex-structure moduli, while  $F_{\Lambda} = \partial F/\partial X^{\Lambda}$ determined by the differential of so-called prepotential ${F}={1\over 6} k_{ijk} X^i X^j X^k/X^0$(in leading order\footnote{In the later discussion, we will discuss on the prepotential with higher order terms which involves more complicated terms mirror to the quantum product of $\hat{X}$.}.) where $k_{ijk}$ is the triple intersection number.

The basis for the even cohomology of Calabi-Yau threefold $X$  can be denoted by $(1,1)$- and $(2,2)$-cohomology, as
\eq{
  \label{res_033a}
  \arraycolsep2pt
  \begin{array}{ccl}
  \{ \omega_{\lab A} \}  &\in&  H^{1,1}(X) \,, \\[5pt]
 \{ \sigma^{\lab A} \} & \in & H^{2,2}(X) \,,
  \end{array}
  \hspace{50pt}\text{with}~ \lab A = 1,\ldots, h^{1,1} \,.
}
Furthermore, one can incorporate  additional six-form $\omega_0= \tfrac{\sqrt{g}}{{\cal V}} \op dx^6$ and zero-form $\sigma^0=1$ to these bases, such that 
\eq{
  \label{res_033}
  \arraycolsep2pt
  \begin{array}{ccl}
  \{ \omega_{ A} \}  &=&  \displaystyle \bigl\{ \tfrac{\sqrt{g}}{{\cal V}} \op dx^6,\, \omega_{\lab A} \bigr\}\,, \\[6pt]
  \{ \sigma^{ A} \} &=& \bigl\{ 1,\,\sigma^{\lab A}\bigr\} \,,
  \end{array}
  \hspace{50pt}A = 0,\ldots, h^{1,1} \,,
}
where $\mathcal V = \int_{X} \sqrt{g} \op d^6x$ denotes the Calabi-Yau threefold volume, and the basis takes the range $0,\ldots, h^{1,1}$ instead. 
The pairing of these bases lead to
\eq{
  \int_{X} \omega_{A}\wedge \sigma^{ B} = \delta_{ A}{}^{ B} \,.
}
The triple intersection of the basis \eqref{res_033a} lead to 
\eq{
  \label{res_025}
  \kappa_{{ijk}} = \int_{X} \omega_{i}\wedge \omega_{j}
  \wedge \omega_{k} \,,
}
while the total volume of the Calabi-Yau threefold in the units $l_s = 2\pi \sqrt{\alpha'}$ can be expressed in
\be
\label{volume}
\mc{V} = \int_{X} J^3 = {1\over 6} \kappa_{ijk} t^i t^j t^k.
\ee
Here $J$ represents the K\"ahler form, while $t^i$, $i=1, \ldots ,h^{1,1}$ are the two-cycle volumes of the Calabi-Yau geometry. 
In the basis $\{\omega_{A}\}$, the K\"ahler form $J$ and the Kalb-Ramond $B$-field can be expressed in terms of 
\eq{
J=t^{ A} \op \omega_{ A} \,, \hspace{70pt}
B = b^{ A}\op\omega_{ A}\,,
}
and contribute to a complexified K\"ahler field as
\eq{
  \mathcal J = J + i B = \bigl( \op t^{ A} + i b^{ A} \bigr)\op \omega_{ A}
  = \mathcal J^{ A} \omega_{ A}\,.
}
In the following study, we denote the complexified K\"ahler moduli by
 \be 
 \label{kahler}
 T_i \equiv \tau_i +i b_i, 
 \ee 
where the axionic fields $b_i$ arise from RR 4-form $C_4$ along four-cycle $D_i$ with $\int_{D_i} C_4 =b_i$, and 
the corresponding four-cycle moduli $\tau_i$ as volume of the four-cycle divisor  $D_i\in H_4(M,\mbb{Z})$, 
\eq{
\label{fourcyc}
\tau_i = \partial_{t_i} {\cal V} ={1\over 2}\int_{X} D_i \wedge J \wedge J = {1\over 2} \kappa_{ijk} t^j t^k.
}
Note that $D_i\in H^2(M,\mbb{Z}$) under Poincare duality.
This allows us to study the total volume in terms of four-cycle moduli $\tau_i$ via~\eqref{fourcyc}.

\subsection{Gukov-Vafa-Witten Superpotential}

In type IIB string compactification, Calabi-Yau orientifold compactification are often utilized to obtain four-dimensional $\mc{N}=1$ theories.
The ten-dimensional bosonic massless fields contains the anti-symmetric self-dual field strength $C_{2}$, $C_{4}$,  NS antisymmetric tensor $B_2$, and metric $g_{MN}$, scalars $\phi, C_{0}$. 
In Calabi-Yau orientifold compactification,  fluxes such as the RR three-form $\mathfrak F_3= d C_2$ and NS three-form $H_3= dB_2$ can be turned on as background fluxes under the  quantisation constraints
\be
\label{quantisation}
\frac{1}{(2 \pi)^2 \alpha'} \int_{\Sigma_a} \mathfrak F_3 = n_a \in \mbb{Z}, \quad \quad \frac{1}{(2 \pi)^2 \alpha'}
\int_{\Sigma_b} H_3 = m_b \in \mbb{Z},
\ee
in which $\Sigma_{a,b}$ are three-cycles of the chosen Calabi-Yau manifolds where the integration performed on. 
Moreover, the generically allowed $O3/O7$ orientifold planes, D3/D7 branes and fluxes are required to satisfy $C_4$ tadpole cancellation condition, such that
\be
\label{tadpolecancellation}
N_{D3} - N_{\bar{D}3} +  \frac{1}{(2\pi)^4 \alpha'^2} \int H_3 \wedge \mathfrak F_3 = \frac{\chi(X)}{24},
\ee
where $\chi(X)$ represents the Euler number of the Calabi-Yau threefold $X$.
The gauge theories arise from the world-volume of both D3 and D7 branes contribute to the model building of standard model or hidden sector matter.
The flux induced superpotential, {\emph{i.e.,}} the Gukov-Vafa-Witten(GVW) superpotential of $\mc{N}=1$ supergravity theory~\cite{Gukov:1999ya} results in
\be
\label{GVW}
W = \int_X G_3 \wedge \Omega,
\ee
where $\Omega$ denotes the holomorphic $\left(3,0\right)$ form, $G_3 = \mathfrak F_3 - i S H_3$ contains the contribution of both the RR and NS three-form and the chiral axio-dilaton denoted by $S=e^{-\phi} - i C_0$.
Here the superpotential depends not only on the dilaton, but also the complex structure
moduli measuring the size of three-cycles where quantisation constraints \eqref{quantisation} performed on. 
Note that the K\"ahler moduli do not enter into the flux induced GVW superpotential.

\subsubsection*{Scalar Potential}
The K\"ahler potential $\mc K$ contains the contributions from all the three moduli: axion-dilaton, complex structure, and the K\"ahler moduli 
$\mc{K}= \mc{K}_{T} + \mc{K}_{U} + \mc{K}_{S}$
taking the form of
\be
\mc{K} =
-2 \ln \left[\mc{V} \right] -
\ln\left[-i \int_M \Omega \wedge \bar{\Omega}\right] - \ln\left(S + \bar{S}\right),
\ee
in which $\cal V$ denotes the total volume of the Calabi-Yau manifold in the units $l_s = 2
\pi \sqrt{\alpha'}$.
The K\"ahler potential $\K_{T}$
 of no-scale type are constraint with $G^{i\bar{\jmath}} \K_i \K_{\bar{\jmath}} = 3$, denoting the K\"ahler metric by
$G_{i\bar{\jmath}} \equiv \mc{K}_{i\bar{\jmath}}=
\partial^2 \mc{K}/\partial_i  \partial_{\bar{\jmath}} $ and  $G^{i\bar{\jmath}}=G^{-1}_{i\bar{\jmath}}$.

The $\mc{N}=1$ supergravity F-term scalar potential, resulting from the flux induced superpotential and K\"ahler potential, takes the form of
\be
\label{Vscalar}
V =  e^{\K} \left[G^{i \bar{j}} D_i W {D}_{\jmath} \bar{W} - 3 \vert W \vert^2 \right],
\ee
where $i,j$ running over all moduli, and $D_i W = \partial_i W + (\partial_i K) W,~ D_i \bar{W} = \partial_i \bar{W} + (\partial_i K) \bar{W}$. The no-scale scalar potential reduces to  
\be\label{Vnoscale}
V_{no-scale} = e^{\K}~ G^{a\bar b}D_a W {D}_b \bar{W},
\ee
while $a$ and $b$ depend on the dilaton and complex-structure moduli only under the no-scale constraints.
The minimum of the $V_{no-scale}$ can be solved by 
\be
D_a W \equiv \partial_a W + (\partial_a K) W = 0,
\ee
with the complex structure and the dilaton moduli stabilized by the introduced background fluxes. 
The value of flux stabilized superpotential at the vacuum can be denoted as $W_0$.

\subsubsection*{Prepotential}

Recall that the K\"ahler potential can be represented in terms of the prepotential by
\eq{\mc{K} = -\ln[X^i \bar F_i(\bar X) + \bar X^i F_i(X)]\,.}
The K\"ahler deformations can be purely captured by prepotential perturbatively. 
The prepotential obtained via mirror symmetry, at large complex structure region, takes a general form of~\cite{Morrison:1991cd, Hosono:1994av, Hosono:1994ax}
\eq{
  \label{PrePotential}
  F = {1\over 6}\kappa_{ijk} X^i X^j X^k/X^0 + \frac{1}{2}~ a_{ij} X^i X^j + b_i X^i X^0 + \frac{1}{2} i \gamma \bigl( X^0\bigr)^2
  + F_{\rm inst.}\,,
}
where $\kappa_{ijk}$ denote the triple intersection numbers of the mirror dual Calabi-Yau manifold $\hat X$ with $\kappa_{ijk}=\int_{\hat X} J_i \wedge J_i \wedge J_k$, while the constant prefactor $a_{ij}$ and $b_i$ are rational numbers 
while $\gamma$ being real with $\gamma=\frac{\zeta(3)~\chi(\hat X)}{(2\pi)^3}$. 
The rational factor $a_{ij}= \frac{1}{2} \int_{\hat X} \iota _*(c_1(J_j)) \wedge J_i~\textit{mode}~\mathbb{Z}$, and $b_{j}=\frac{1}{4!} \int_{\hat X} c_2(\hat X) \wedge J_j$.
In which, $c_1(J_j)$ denotes the first Chern class of the divisor associated to $J_j$ and $\iota _*(c_1(J_j))=P_{\hat{X}} i_* P^{-1}_{J_j}$ is the Gysin homomorphism with $i_*$ as the push-forward on the homology, and $P_{\hat{X}}$ denoting the Poincare-duality map. In total $\iota _*(c_1(J_j))$ become a four-form \cite{Grimm:2009ef}.
Here,  $\chi(\hat X)$ is the Euler characteristic of the $\hat X$, and $J_i\in H^{1,1}(\hat X,\mathbb Z)$ is the basis of harmonic $(1,1)$-forms. 

Note that the ansatz of the prepotential~\eqref{PrePotential} is valid limited to the large complex structure region that is convergence~\cite{Hosono:1994av}. The radius of convergent region shall be determined by the singularity of its associated Yukawa couplings given in~\cite{Candelas:1994hw,Klemm:1999gm}. 
In practice, this can be taken as a constraint with a given cutoff according to finite Gopakumar–Vafa invariants, such that
\begin{equation}\label{eq:InstCutoff}
    \biggl |\dfrac{F_{\text{inst}}}{F_{\text{pert}}}\biggl |<\varepsilon,
\end{equation}
in which $F_{\text{inst}}$ is often computed up to a finite cutoff according to the value of $\varepsilon$~\cite{Dubey:2023dvu, Demirtas:2020ffz,Alvarez-Garcia:2020pxd}.
The total perturbative corrections to the ansatz prepotential can be incorporated as a rational shift to the  background fluxes.

In the usual case, the mirror dual manifold $\hat{X}$ is different manifold than the original Calabi-Yau manifold $X$. To arrive in the large complex structure and large volume limit in the meanwhile for one specific Calabi-Yau compactification,  special self-mirror Calabi-Yau manifolds with $X=\hat{X}$ can be utilized. 
For which, the Euler characteristic $\chi(\hat X)=0$, therefore the $\alpha'$ correction to the effective scalar potential in large volume limit became trivial and leads to a special large volume scenario.

\section{Self-mirror Calabi-Yau Geometry}

In this section, we investigate in the geometry of self-mirror Schoen Calabi-Yau manifolds with the total volume explicitly given, and then study the self-mirror large volume scenario therein.
Firstly, take Schoen Calabi-Yau threefold as example, we study the special type of self-mirror Calabi-Yau geometry which was utilized in string theory, such as in \cite{Curio:1997rn, Curio:1998bv, Deffayet:2024hug}. 

\subsection{Schoen Calabi-Yau Threefold as a double fibration}

From complete intersections, 
Schoen type Calabi-Yau threefold can be constructed as the fiber products
of two del Pezzo $d P_9$ surfaces, $B_1$ and $B_2$, which are fibered over $\mathbb P^1$. In which, the $d P_9$ surface can be defined as a blow-up of $\mathbb P^2$ at $9$ points where all the $9$ points are distinct with no Kodaira fibers collide~\cite{Hosono1997mirror, Braun:2007sn, Schoen:1988, Braun:2004xv, Donagi:2000zf, Braun:2005nv, Braun:2005ux, Ovrut:2002jk, Braun:2005zv, Braun:2007xh, Braun:2007vy, Braun:2007tp}.
First, recall that the ambient variety
$\mathbb P^2 \times \mathbb P^1 \times \mathbb P^2$ with coordinates $x_i, t_j, y_k$, we have
\begin{equation}
\Big( [x_0:x_1:x_2],~ [t_0:t_1],~[y_0:y_1:y_2]\Big) \in 
\mathbb P^2 \times \mathbb P^1 \times \mathbb P^2.
\end{equation}
The Schoen Calabi-Yau threefold can be defined as the zero-set of two equations 
\begin{equation}
	\label{eq:P}
	\tilde P(x,t,y) ~=~ 
	t_0 \tilde P_1\big(x_0, x_1, x_2\big) + 
	t_1 \tilde P_2\big(x_0, x_1, x_2\big)
	= 0 
	,
	\end{equation}
	\begin{equation}
	\label{eq:R}
	\tilde R(x,t,y) ~=~ 
	t_1 \tilde R_1\big(y_0, y_1, y_2\big) + 
	t_0 \tilde R_2\big(y_0, y_1, y_2\big)  
	= 0 
\end{equation}
with multi-degrees $(3,1,0)$ and $(0,1,3)$, respectively. 
Therefore, the Schoen Calabi-Yau threefold $X$ is constructed as a complete intersection, while the 
ambient space $\mathbb P^2 \times \mathbb P^1 \times \mathbb P^2$ is a toric variety.
The first Chern class of line bundles on $X$ form a lattice of dimension $h^{1,1}\big(\tilde X \big)=19$.
Among which, the ambient space with $h^{1,1}\big(\mathbb P^2\times\mathbb P^1\times\mathbb P^2\big)=3$ provides $3$ number of ambient divisors and the rest $16$ divisors come from the non-ambient space with associated line bundles $\cal L(D)$ that are not toric
\footnote{
Note that an alternative way to embed Schoen Calabi-Yau threefold in a more complicated toric variety where all divisors are toric was discussed in~\cite{Braun:2007vy}.}.
Again the Schoen threefold can be constructed as  
\be \label{CY19}
CY^{(19,19)}={\tiny 
	\left[\begin{array}{c|cc}P^2&3&0\\P^1&1&1\\P^2&0&3\end{array}\right]}=
dP\times _{P^1_y}dP,
\ee
elliptically fibered over del Pezzo $d P_9$ surfaces, and can also be constructed  from
$T^2 \times K3={\tiny 
	\left[\begin{array}{c|cc}P^2&3&0\\P^1&0&2\\P^2&0&3\end{array}\right]}$
via Voisin-Borcea involution\cite{Curio:1997rn}.
The Hodge numbers of Schoen Calabi-Yau threefold constituting the self-mirror hodge diamond, such that
\begin{equation}
h^{p,q}\big(\tilde X\big) = ~
\vcenter{\xymatrix@!0@=7mm@ur{
		1 &  0  &  0  & 1 \\
		0 &  19 &  19 & 0 \\
		0 &  19 &  19 & 0 \\
		1 &  0  &  0  & 1 
}}.
\end{equation}
Note that, alternatively, one can bypass the non-toric difficulties by investigating a certain $\mathbb Z_3 \times \mathbb Z_3$-quotient of Schoen Calabi-Yau threefold, for which only the toric line bundles are relevant~\cite{Braun:2007sn}.

\subsection{Quotient Schoen Calabi-Yau Threefold}

\begin{figure}[ht]\centering
	\subfigure[]{\includegraphics[width=0.45\linewidth]{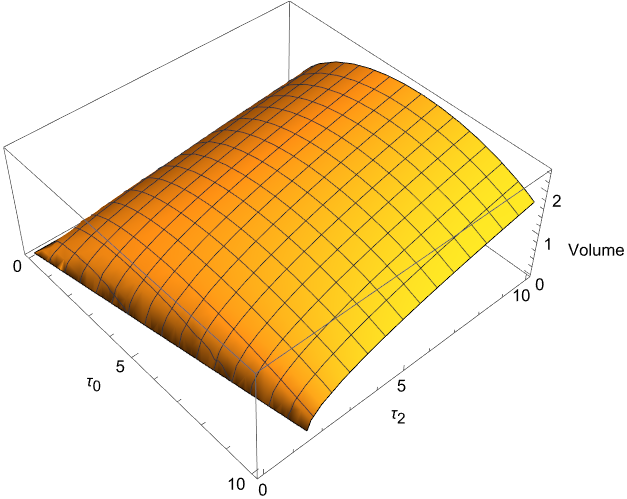}}\qquad
	\subfigure[]{\includegraphics[width=0.45\linewidth]{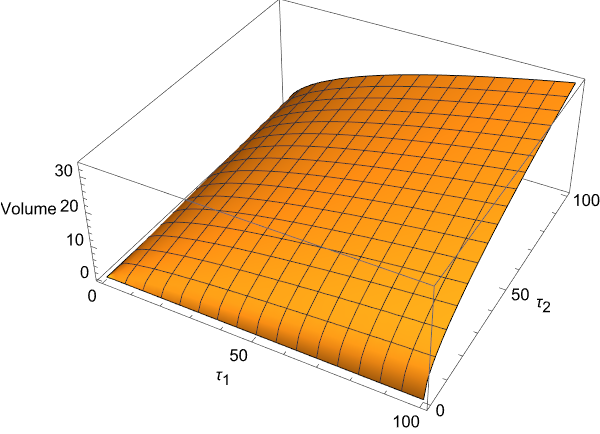}}
	\caption{Total volume in K\"ahler moduli space according to $\tau_0, \tau_1$ and $\tau_2$.}
\label{figure:Qvolume-t12}
\end{figure}

As mentioned in~\cite{Braun:2007sn}, 
the quotient version of Schoen Calabi-Yau threefold inherits the double elliptic fibration properties of Schoen Calabi-Yau threefold yet only with three ``ambient'' moduli.
As follows, the $\mathbb Z_3 \times \mathbb Z_3$-fixed points are quotient out from the Schoen's Calabi-Yau threefold with the quotient action
 \begin{equation}
 X = \tilde X \Big/ \big(\mathbb Z_3 \times \mathbb Z_3\big) 
 = \Big\{ \tilde P=0=\tilde R \Big\} \Big/ \big(\mathbb Z_3 \times \mathbb Z_3\big).
 \end{equation}
This results in a smooth Calabi-Yau threefold with fundamental group
 $\pi_1(X)=\mathbb Z_3 \times \mathbb Z_3$ as quotient Scheon Calabi-Yau threefold whose Hodge diamond given by~\cite{Braun:2004xv} 
 \begin{equation}
 \label{eq:SchoenZ3Z3Hodge}
 h^{p,q}\big( X \big)
 =h^{p,q}\big( \tilde X \Big/ \big(\mathbb Z_3 \times \mathbb Z_3\big)\big)=
 \vcenter{\xymatrix@!0@=7mm@ur{
 		1 &  0 &  0 & 1 \\
 		0 & 3 & 3 & 0 \\
 		0 & 3 & 3 & 0 \\
 		1 &  0 &  0 & 1 
 }}
 \,.
 \end{equation}
In such a way, a quotient version of Schoen Calabi-Yau threefold is constructed representing the 
double elliptic fibration yet in fully ambient space. 
From the ambient space $\mathbb{P}^{2}\times \mathbb{P}^{1}\times \mathbb{P}^{2}$,
the class of the variety $X$ itself as a complete intersection is
\begin{equation}
[X]=(3H_1+H)(3H_{2}+H)\in A^{2}(\mathbb{P}^{2}\times \mathbb{P}^{1}\times \mathbb{P}^{2})\,,
\end{equation}
where $A^{*}$ is the integral Chow ring,
and $H^{2}=0\in A^{2}(\mathbb{P}^{2}\times \mathbb{P}^{1}\times \mathbb{P}^{2})$. It follows that curve classes are given by linear combinations of
\begin{equation}\label{c2element}
H_{1}|_{X}^{2} \,,
\quad
H_{1}|_{X}H|_{X}\,,
\quad
H_{2}|_{X}^{2}\,,
\quad
H_{2}|_{X}H|_{X}\,,
\quad
H_{1}|_{X}H_{2}|_{X}
 \,,
\end{equation}
where $|_{X}$ stands for restriction to $X$.
Using the above formula for the class $[X]$ and properties of  $A^{2}(\mathbb{P}^{2}\times \mathbb{P}^{1}\times \mathbb{P}^{2})$,
it follows that inside $X$, the nontrivial triple intersections of Schoen Calabi-Yau threefold in the ambient space are
\begin{equation}\label{eqnambientcohomologyintersection}
H_{1}^{2}H_{2}[X]=3\,,
\quad
H_{2}^{2}H_{1}[X]=3\,,
\quad
H_{1}HH_{2}[X]=9\,.
\end{equation}
Then, the total volume of quotient Schoen Calabi-Yau threefold takes the form only involving the ambient space, such that the total volume can be represented by
\bea\label{qvolume}
{\cal V}
&=&{3\over 2}\,
 t_{0}^{(1)}t_{0}^{(1)}t_{0}^{(2)}
 + {3\over 2} \,t_{0}^{(1)}t_{0}^{(2)}t_{0}^{(2)}
 +9\, t^{(0)} t_{0}^{(1)}t_{0}^{(2)}
.
\eea
The corresponding four-cycle moduli $\tau_i$ can be derived accordingly  with 
\bea
&\tau_1 = \partial_{t_{0}^{(1)}} {\cal V} = {3\over 2}\, (t_{0}^{(2)})^2+ {3\over 2}\,t_{0}^{(1)} t_{0}^{(2)} + 9\, t^{(0)} t_{0}^{(2)} ,\\ \nonumber
&\tau_2 = \partial_{t_{0}^{(2)}} {\cal V} = {3\over 2}\, (t_{0}^{(1)})^2+ {3\over 2}\,t_{0}^{(1)} t_{0}^{(2)} + 9\, t^{(0)} t_{0}^{(1)} ,\\ \nonumber
&\tau_s = \partial_{t^{(0)}} {\cal V} = 9\, t_{0}^{(1)} t_{0}^{(2)} .
\eea
As follows, the two-cycle volumes can be written into
\bea
&t_{0}^{(1)} = \frac{\sqrt{\tau_0} \sqrt{6 \tau_2-\tau_0}}{3 \sqrt{6 \tau_1-\tau_0}},\\ \nonumber
&t_{0}^{(2)} = \frac{\sqrt{\tau_0} \sqrt{6 \tau_1-\tau_0}}{\sqrt{54 \tau_2-9 \tau_0}},\\ \nonumber
&t^{(0)} = \frac{-4 \tau_0 (\tau_1+\tau_2)+12 \tau_1 \tau_2+\tau_0^2}{6 \sqrt{\tau_0} \sqrt{6 \tau_1-\tau_0} \sqrt{6 \tau_2-\tau_0}},
\eea
and the total volume can be expressed with the k\"ahler moduli, such that
\be\label{vQSchoen}
  {\cal V}= {1\over 18}\sqrt{\tau_0 (\tau_0-6 \tau_1)(\tau_0-6 \tau_2)}.
\ee
The total volume of quotient Schoen Calabi-Yau threefold can be illustrated in Figure \ref{figure:Qvolume-t12} as below with finite value of fiber divisor $\tau_1=2$ and base divisor $\tau_0=1$, respectively\footnote{Since in the later large volume scenario study it only involves with the region with $\tau_i>0$, we only plot with $\tau_i>0$.}. The z-axis represents the value behavior of the total volume.
According to one single ambient fiber divisor $\tau_1$ or $\tau_2$, a bounded behavior shows. However, with both $\tau_1$ and $\tau_2$ goes to large, the total volume is able to reach the large volume limit with finite value of base divisor $\tau_0$. Also, a symmetric shape of the volume appears due to the double elliptic construction of Schoen Calabi-Yau manifold.

\subsection{Schoen Calabi-Yau Volume with Non-ambient}

\begin{figure}[ht]\centering
	\subfigure[]{\includegraphics[width=0.43\linewidth]{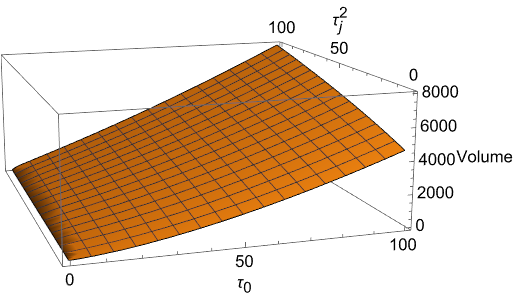}}\qquad
	\subfigure[]{\includegraphics[width=0.43\linewidth]{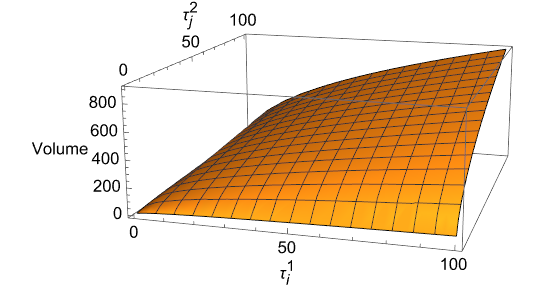}}
	\caption{Total volume in K\"ahler moduli space according to $\tau_0, \tau_i^{(1)}$ and $\tau_j^{(2)}$.}
\label{figure:Svolume-t12}
\end{figure}

To study the volume form of Schoen Calabi-Yau threefold, besides the ambient space discussed above, one needs to incorporate the non-ambient space as well. 
Follow the double elliptic fibration construction of Schoen Calabi-Yau \cite{Hosono1997mirror}, the fiber product from two copies of elliptic fibrations
$\pi_a: S_{a}\rightarrow \mathbb{P}^{1},~  a=1,2$ induce the  Schoen threefold 
\begin{equation}
\pi:
X=S_1\times_{\mathbb{P}^{1}} S_{2}
\rightarrow \mathbb{P}^{1}\,
\end{equation}
with the induced morphisms $p_{a}: X \rightarrow S_{a},~ a=1,2$.
Besides the ambient triple intersection given in \eqref{eqnambientcohomologyintersection}, 
the triple intersection involving with non-ambient divisors are  
\[
H\cdot E_{i}^{(a)}\cdot E_{j}^{(b)}\,,\quad
 E_{i}^{(a)}\cdot E_{j}^{(b)}\cdot E_{k}^{(c)}\,,\quad a,b,c\in\{1,2\}\,,\quad i,j,k\in \{0,1,2\cdots, 8\}\,,
 \]
where $E_{i}^{(a)}, E_{j}^{(b)}$ are the elliptic fibration divisors from both ambient and non-ambient spaces. In which, we have 
\begin{equation}    E_{i}^{(a)}E_{j}^{(a)}=0\,\quad \text{if}~i\neq j\,,
\end{equation}
as $e_{i}^{(a)}\cdot e_{j}^{(a)}=\delta_{ij}$  in $\mathrm{Pic}(S)$.
Therefore the only non-vanishing triple intersection are those of the form (now with $a\neq b$)
\[
H\cdot E_{i}^{(a)}\cdot E_{i}^{(a)}\,,
\quad 
H\cdot E_{i}^{(a)}\cdot E_{j}^{(b)}\,,\quad 
E_{i}^{(a)}\cdot E_{i}^{(a)}\cdot E_{i}^{(a)}\,,\quad
E_{i}^{(a)}\cdot E_{i}^{(a)}\cdot E_{j}^{(b)}\,.
\]
Follow the double fibration construction \cite{Hosono1997mirror}, to incorporate the non-ambient intersection, we introduce the relation of ambient and non-ambient classes 
\begin{equation}\label{eqnbasiswithHa}
 3H_{a}-\sum_{i=0}^{8} E_{i}^{(a)}
 =p_{a}^{*}(f_{a})=H\,,\quad a=1,2.
 \end{equation}
The non-trivial triple intersections can be computed through intersection theory with\footnote{Here we acknowledge Jie Zhou for helping with deriving the triple intersection from algebraic geometry and intersection theory methods.} 
\begin{equation}\label{eqntripleintersections}
H\cdot E_{i}^{(a)}\cdot E_{j}^{(b)}=
H\cdot e=1\,,
\quad
E_{i}^{(a)}\cdot E_{i}^{(a)}\cdot E_{j}^{(b)}=(e\cdot e)_{S_{b}}
=-1\,.
\end{equation}
As a consistency check, \eqref{eqnambientcohomologyintersection} can be re-derived using the above triple intersection results and the relation of ambient and non-ambient classes induced in \eqref{eqnbasiswithHa}.
The non-trivial triple intersections in \eqref{eqntripleintersections} can be summarized as a generating series as follows.
Let 
\eq{\label{abasis}
\omega=s^{(0)}H+\sum_{i=0}^{8}s_{i}^{(1)}E_{i}^{(1)}+\sum_{i=0}^{8}s_{i}^{(2)}E_{i}^{(2)}
}
be an element in $H^{1,1}$.
Utilizing the basis $\{H, E_{i}^{(1)},E_{i}^{(2)},i=0, 1,2,\cdots 8\}$,
the total volume of the Calabi-Yau~\eqref{volume} can be represented by
two-cycle volumes $s^{(0)}, s_i^{(1)}, s_j^{(2)}$, in explicit as
\eq{\label{svolume}
\mathcal{V}={1\over 6} \omega^3
=\sum_{i,j=0}^{8} s^{(0)} s_{i}^{(1)}s_{j}^{(2)}
-{1\over 2}\sum_{i,j=0}^{8}  (s_{i}^{(1)}s_{i}^{(1)} s_{j}^{(2)}+s_{i}^{(1)} s_{j}^{(2)}s_{j}^{(2)})\,,
}
with permutation of two-cycle volumes considered. 
The corresponding four-cycle moduli $\tau_i$ can be derived accordingly with 
\bea
&\tau_{i}^{(1)} = \partial_{s_{i}^{(1)}} {\cal V} =  s^{(0)} s_{i}^{(2)}-s_{i}^{(1)}s_{j}^{(2)} - \frac{1}{2}\, s_{j}^{(2)}s_{j}^{(2)} ,\\ \nonumber
&\tau_{j}^{(2)} = \partial_{s_{j}^{(2)}} {\cal V} =  s^{(0)} s_{i}^{(1)}-s_{i}^{(1)}s_{j}^{(2)} - \frac{1}{2}\, s_{i}^{(1)}s_{i}^{(1)} ,\\ \nonumber
&\tau_0 = \partial_{s^{(0)}} {\cal V} =  s_{i}^{(1)} s_{j}^{(2)} .
\eea
As follows, the two-cycle volumes can then be written into
\bea
&s_{i}^{(1)} = \frac{\sqrt{\tau_0(\tau_0 +2 \tau_{j}^{(2)}) }}{\sqrt{\tau_0 +2 \tau_{i}^{(1)} }},\\ \nonumber
&s_{j}^{(2)} = \frac{\sqrt{\tau_0(\tau_0 +2 \tau_{i}^{(1)}) }}{\sqrt{\tau_0+2 \tau_{j}^{(2)} }},\\ \nonumber
&s^{(0)} =  \frac{3\tau_0^2
+4(\tau_0 \tau_{i}^{(1)}+\tau_0 \tau_{j}^{(2)}+\tau_{i}^{(1)} \tau_{j}^{(2)})}{2 \sqrt{\tau_0(\tau_0 +2 \tau_{i}^{(1)})(\tau_0+2 \tau_{j}^{(2)}) }},
\eea
and the total volume can therefore be expressed with the k\"ahler moduli $\tau_i$, such that
\be\label{vSchoen-s}
  {\cal V}= {1\over 2}\sum_{i,j=0}^{8}\sqrt{\tau_0 (\tau_0+2 \tau_{i}^{(1)})(\tau_0+2 \tau_{j}^{(2)})}.
\ee
The total volume of Schoen Calabi-Yau threefold can be illustrated in Figure \ref{figure:Svolume-t12} as below with 
finite value of fiber divisor $\tau_i^{(1)}=2$ and base divisor $\tau_0=1$, respectively. The z-axis represents the value behavior of the total volume.
According to the fiber divisor $\tau_i^{(1)}$ or $\tau_j^{(2)}$, combination of ambient and non-ambient divisors, a bounded behavior still shows as we observed in the quotient Schoen Calabi-Yau case as in Figure \ref{figure:Qvolume-t12}. 

To reach the large volume limit, we have two approaches. Firstly, due to the bounded behavior of the total volume according to the fiber divisors $\tau_i^{(1)}, \tau_j^{(2)}$ with finite $\tau_0$, to reach the large volume limit $\tau_i^{(1)}, \tau_j^{(2)}$ shall be exponentially large at the meanwhile. 
Alternatively, 
we can set the base divisor $\tau_0$ to be exponentially large while the fiber divisors $\tau_i^{(1)}$ and $\tau_j^{(2)}$ contributes non-perturbatively.

\section{Self-mirror Large Volume Scenario}

Being able to reach the large volume limit in K\"ahler moduli space properly for quotient and  Schoen Calabi-Yau threefold, we are ready to study the self-mirror large volume scenario. 
The total volume of quotient and Schoen Calabi-Yau threefold were given in \eqref{vQSchoen} and  \eqref{vSchoen-s} in the units of $l_s = 2\pi \sqrt{\alpha'}$, in which the triple intersections take a crucial role. 

Follow the standard large volume scenario with $h^{1,2}>h^{1,1}>1$ \cite{Balasubramanian:2005zx, Conlon:2005ki}, here we study the higher order corrections at the large volume limit.
This involves the perturbative $\alpha'$ and non-perturbative contributions to the $\mc{N}=1$ supergravity scalar potential ~\eqref{Vscalar}.

\subsection{Higher Order Type IIB Scalar Potential}

For quotient and Schoen Calabi-Yau threefold, with the total volume derived in  \eqref{vQSchoen} and \eqref{vSchoen-s}, here we study the higher order corrections with {\emph{trivial} Euler characteristic $\chi$ from self-mirror construction.
Distinct with the general study in~\cite{Balasubramanian:2005zx, Conlon:2005ki}, here for self-mirror Schoen threefold $\chi = 0$ with $h^{2,1} = h^{1,1}$, \emph{i.e.}, equal number of complex structure and K\"ahler moduli. In this circumstance, 
the standard large volume scenario dramatically change in $\alpha'$ correction.
let's first recall the basis of leading order correction in $\mathcal{O}(\alpha'^3)$ and then study the self-mirror large volume scenario from Schoen compactification. 

For self-mirror Calabi-Yau manifolds,  the  prepotential  at large complex structure limit via mirror symmetry will be different than \eqref{PrePotential}.
Since the Euler characteristic $\chi(\hat X)=0$, the imaginary term trivially annihilates, and the prepotential \eqref{PrePotential} reduces to
\be \label{Fself}
    F = {1\over 6}\kappa_{ijk} X^i X^j X^k/X^0 + \frac{1}{2} a_{ij} X^i X^j + b_i X^i + F_{\rm inst.}\,,
\ee
with $a_{ij}= \frac{1}{2} \int_{\hat X} \iota _*(c_1(J_j)) \wedge J_i~\textit{mode}~\mathbb{Z}$, $b_{j}=\frac{1}{4!} \int_{\hat X} c_2(\hat X) \wedge J_j$ and $c_2(\hat X)$ denotes the second Chern.
The $\alpha'$-correction in larege volume scenario of type IIB string theory corresponds to the imaginary contribution $i \gamma \bigl( X^0\bigr)^2$ with $\gamma=\frac{\zeta(3)~\chi(\hat X)}{(2\pi)^3}$ vanishes due to vanishing Euler characteristic $\chi(\hat X)=0$ \cite{Balasubramanian:2005zx, Conlon:2005ki}.  
Therefore, the $\alpha'$-correction vanish trivially as well.

Apart from the $\alpha'$-correction to the K\"ahler potential, the instantons and/or gaugino condensation contribute non-perturbatively to the superpotential as well \cite{Witten:1996bn}. 
These contributions get incorporated into the effective scalar potential as in \eqref{Vscalar}.
For a general Calabi-Yau manifold, 
the K\"ahler potential 
in the units of $l_s = 2\pi \sqrt{\alpha'}$ is given in~\cite{Becker:2002nn, Grimm:2004uq}
 \be \label{KahlerPotential} 
\mc{K} = -2 \ln
\left[ \mathcal{V} + \frac{\xi}{2 g_s^{3/2}}\right] - \ln\left[-i \int_M
\Omega \wedge \bar{\Omega}\right] - \ln\left[\left(S+\bar{S}\right)\right], 
\ee 
in which the term with $\xi = - \frac{\zeta\left(3\right) \chi\left(M\right)}{2(2 \pi)^3}$ vanishes in self-mirror compactification. 
With the dilaton and complex structure moduli dependences integrated out to $K_{cs}$, the K\"ahler potential reduces to 
\be \label{Ksim}
\mc{K}  =  \mc{K}_{cs}
-2 \ln \left[ \mc{V} +
\frac{\xi}{2 g_s^{3/2}} \right] \xrightarrow{\xi=0} \mc{K}_{cs}-2 \ln \mc{V} ,
\ee
in which the term factored with $\xi = - \frac{\zeta\left(3\right) \chi\left(X\right)}{2(2 \pi)^3}$ vanishes again for self-mirror Calabi-Yau manifolds. 
The superpotential contributes to the scalar potential non-perturbatively with D3-brane instantons and/or gaugino condensation when wrapped D7-branes introduced \cite{Witten:1996bn}, such that 
\be
\label{supercorrect}
W = \int_M G_3 \wedge \Omega + \sum_i A_i e^{- a_i T_i},
\ee
where $T_i \equiv \tau_i +i b_i$, $A_i$ represents the one-loop determinant which depends on the complex structure moduli for D3-brane instantons case.
 $a_i = \frac{2 \pi}{K}$ with $K=1$ for D3-instantons case and $K \in \mbb{Z}_+$ for general cases. 
By integrating out the dilaton and complex structure moduli dependence to $W_0$, the superpotential with non-perturbative corrections can be represented by
\be\label{Ssim}
W  =  W_0 + \sum_i A_i e^{- a_i T_i}.
\ee
The effective scalar potential~\eqref{Vscalar} 
implemented with K\"ahler potential~\eqref{Ksim} and superpotential~\eqref{Ssim}, results in \cite{Balasubramanian:2004uy, Conlon:2005ki}
\bea
\label{potential}
V 
& = & V_{np1} + V_{np2} + V_{\alpha'} \\ \nonumber
& = & e^\mc{K} \left[ G^{i \bar{j}} \left(a_i A_i a_j \bar{A}_j e^{- \left(a_i T_i + a_j \bar{T}_j\right)}
-  ( a_i A_i e^{-a_i T_i} \bar{W} \partial_{\bar{j}} K + a_j \bar{A}_j e^{- a_j \bar{T}_j}
W \partial_{i} K) \right) \right .\nonumber \\
& + & \left. 3 \xi \frac{\left(\xi^2 + 7\xi \mc{V} +
\mc{V}^2\right)}{\left(\mc{V} - \xi\right)\left(2\mc{V} +
\xi\right)^2}
|W|^2\right].
\eea
in which, at the large volume limit $\mc{V} \to \infty$, by incorporating the dilaton-dependent $e^{-\frac{3 \phi_0}{2}}$ into $e^{\mc{K}_{cs}}$, with
$\mc{K} = \mc{K}_0=\mc{K}_{cs}-2 \ln \mc{V}$, 
K\"ahler term $e^\mc{K}$ results in
\be
\label{eK}
e^\mc{K} \sim \frac{e^{\mc{K}_{cs}}}{\mc{V}^2} + \mc{O}\left( \frac{1}{\mc{V}^3} \right).
\ee
Therefore, for self-mirror compactification with $\xi=0$, the non-perturbative terms from k\"ahler moduli contribute to the effective scalar potential \eqref{Vscalar} as\footnote{Note that as the axionic fields $b_i$ can be stabilized to constant coefficients, here we denote $\partial_{{T_i}} \K=\partial_{{\tau_i}} \K:=\partial_{{i}} \K$.} 
\bea \label{VQ}
V & = & {1\over \mc{V}^2} \left[ G^{i \bar{j}} \left(a_i A_i a_j \bar{A}_j e^{- \left(a_i T_i + a_j \bar{T}_j\right)}-  ( a_i A_i e^{-a_i T_i} \bar{W} \partial_{{j}} K + a_j \bar{A}_j e^{- a_j \bar{T}_j} W \partial_{i} K) \right) \right],
\eea
by further incorporating the factor $e^{\mc{K}_{cs}}$ into $A_s$ and $W$. Here it is obvious that the volume form and the triple intersection of the particular Calabi-Yau manifold play an important role in deriving the non-perturbative contributions to the effective scalar potential. 

Moreover, so far we have studied the scalar potential in the K\"ahler moduli space with the dilaton and complex structure integrated out to $e^{\mc{K}_{cs}}$. To reconcile the scalar potential minimum in the whole moduli space, one shall get back dilaton and complex structure moduli dependence, such that the whole scalar potential for self-mirror Calabi-Yau summed up to
\bea \label{Vcsdilaton}
V & = & e^\mc{K} (G^{a\bar{b}} D_a W {D}_b \bar{W} + G^{\tau \bar{\tau}}
D_\tau W {D}_\tau \bar{W}) 
+ V_{np1} + V_{np2},
\eea
in which the first terms factored with $e^K$ incorporating the dependence of dilaton and complex structures. These terms are trivialized at the point $D_\tau W = D_{\phi_i} W = 0$ in the moduli space. 
However, by moving away from this point along dilaton and complex structure directions, the scalar potential shall achieve positive value of $\mc{O}\left(\frac{1}{\mc{V}^2}\right)$
according to \eqref{eK}.  
While this is in the same order or higher order than the non-perturbative terms $V_{np1}$ and $V_{np2}$,
the scalar potential minimum along the K\"ahler direction will indeed be the vacuum for the whole moduli space.

\subsection{Quotient Schoen Calabi-Yau in Explicit}

With explicit choice of flux in type IIB flux compactification, as discussed in \cite{Blumenhagen:2015xpa}, complex structure moduli can be stabilized by solving
\be
\label{StabilisingModuli}
D_\tau W_{cs} = 0 \quad \textrm{ and } \quad D_{\phi_i} W_{cs} = 0~,
\ee
where $W_{cs}$ denotes the flux induced GVW superpotential depending on the dilaton and complex structure moduli. Integrating out these dependence of superpotential to $W_0$, here we study the large volume scenario in the K\"ahler moduli space.

Explicitly for quotient Schoen Calabi-Yau threefold with $h^{1,1}=h^{2,1}=3$, one can solve the total volume into the K\"ahler moduli space with the solution \eqref{qvolume}.
In which, K\"ahler moduli $\tau_1$ and $\tau_2$ are volumes of the divisors $D_1$ and $D_2$ which correspond to the two-cycle volumes $t_0^{(1)}, t_0^{(2)}$ from the double elliptic fibers.
K\"ahler moduli $\tau_0$ is the  volume of base divisor $D_0$ corresponding to the two-cycle volumes $t^{(0)}$ from base $\mathbb{P}^{1}$.
Set $\tau_0, \tau_1, \tau_2$ correspond to the divisors $D_0$,  $D_1$ and $D_2$ which can be embedded into a Calabi-Yau fourfold with elliptic torus fibration over a base threefold.
These contributions are manifested in the superpotential as the non-perturbative terms 
\be\label{W}
W = W_0 + A_0 e^{- a_0 T_0}+ A_1 e^{-a_1 T_1} + A_2 e^{-a_2 T_2},
\ee
in which $A_i$ are the one-loop determinant 
and $a_i = \frac{2 \pi}{K}$ with $K=1$ for D3-instanton case, while $K \in \mbb{Z}_+$ for general cases. 
To obtain the concrete non-perturbative terms, we first compute the inverse of K\"ahler metric for deriving the correction to the  effective scalar potential \eqref{Vscalar}
\eq{\label{Kmetric}
G^{i\bar{j}}=
\begin{pmatrix}
\tau_1^2-\frac{\tau_1 \tau_0}{3}+\frac{\tau_0^2}{18} & {\tau_0^2 \over 36} & \qquad {\tau_0^2 \over 6}\\
\newline\\
{\tau_0^2 \over 36} & \tau_2^2-\frac{\tau_2 \tau_0}{3}+\frac{\tau_0^2}{18} & \qquad {\tau_0^2 \over 6}\\
\newline\\
{\tau_0^2 \over 6} & {\tau_0^2 \over 6} & \qquad \tau_0^2
\end{pmatrix}.
}
It can be verified that $G^{i\bar{j}}\partial_i {\cal K}~\partial_{\bar{j}} {\cal K}= 3$, therefore no-scale structure of scalar potential preserved.

Based on the former large volume limit study, there are two approaches to obtain the non-perturbative corrections to the effective scalar potential. First, set  $\tau_1, \tau_2$  to be exponentially large while the K\"ahler moduli $\tau_0$ with a finite value. Second, set $\tau_0$  to be exponentially large while the K\"ahler moduli $\tau_1, \tau_2$ both with a finite value.

\subsubsection{Large Volume Scenario I}

In the large volume limit with fiber K\"ahler moduli $\tau_1, \tau_2$ both to be exponentially large,
we have briefly shown in \cite{Sun:2024ral} that
the non-perturbative contribution comes from
\be
W = W_0 + A_0 e^{-a_0 T_{0}},
\ee
when the base divisor $\tau_0$ with a finite value and $T_{0}=\tau_0+ i b_0$.
The non-perturbative terms contributes to the effective scalar potential \eqref{VQ}, such that
\be\label{VQSchoen1}
V=\frac{W_0^2 (\tau_0 \tau_1 \tau_2)^2}{81\mc{V}^6} 
-\frac{ 2 W_0^2 \tau_0^3 \tau_1 \tau_2(\tau_1+\tau_2)}{243\mc{V}^{6}}
+\frac{2 A_0 W_0 e^{-a_0 \tau_0} (\tau_0 \tau_1 \tau_2)^2\cos(a_0 b_0)(1+a_0 \tau_0)}{81\mc{V}^{6}},
\ee
with $G^{00}=\tau_0^2$ read off from the inverse of K\"ahler metric \eqref{Kmetric}. 
The total volume of Schoen Calabi-Yau \eqref{vQSchoen} with $\tau_1, \tau_2 \to \infty$ behaves as
\bea
  {\cal V} &=& 
  {1\over 18}\sqrt{\tau_0^3-6 \tau_0^2(\tau_1+\tau_2)+ 36 \tau_0 \tau_1 \tau_2} \\ \nonumber
 &\sim & {1\over 3} \sqrt{ \tau_0\tau_1 \tau_2}, ~~
 \text{with}~ \tau_0 \tau_1 \tau_2 \sim (3 \mc{V})^{2}.
\eea
Consider that the K\"ahler moduli $\tau_1, \tau_2$ taking equal positions in the K\"ahler moduli space, the scalar potential \eqref{VQSchoen1} in the order of $\mathcal{O} (\frac{1}{\mc{V}^{n}})$ 
can be reduced to 
\bea\label{VSchoen001}
V &=& \frac{W_0^2}{\mc{V}^2} 
-\frac{4 W_0^2 \tau_0^{3/2}}{9\mc{V}^{3}}
+\frac{2  A_0 W_0 e^{-a_0 \tau_0}\cos(a_0 b_0)(1+a_0 \tau_0)}{\mc{V}^{2}}.
\eea
The subleading term of the scalar potential with $\cos(a_0 b_0)$ are suppressed according to the axionic field $b_0$. 
Here $a_0 \tau_0 \gg 1$ shall be required to ignore the higher order instanton corrections, and $a_0 \tau_0 \sim \ln \mc{V}$ are usually taken as a decompactification in the large volume limit $\mc{V} \to \infty$.
And therefore the scalar potential in the order $\mathcal{O} (\frac{1}{\mc{V}^{n}})$ reduces to 
\bea \label{P1}
V &=& \frac{W_0^2}{\mc{V}^2} 
-\frac{4 W_0^2 \tau_0^{3/2}}{9\mc{V}^{3}}
+\frac{2  A_0 W_0 \cos(a_0 b_0)(1+\ln \mc{V})}{\mc{V}^{3}}.
\eea
Distinct with the standard large volume scenario while $h^{1,2}>h^{1,1}>1$, the self-mirror non-perturbative corrections obtain a clear uplift term in the order of $\mathcal{O} (\frac{1}{\mc{V}^{2}})$ from the symmetric construction of Schoen Calabi-Yau manifold. 
The scalar potential approaches zero from above as the first term $\frac{W_0^2}{\mc{V}^2}$ dominates as illustrated in Figure \ref{figure:SPotential-QS1} according to \eqref{VQSchoen1}.
\begin{figure}[h]
 	\centering 	\includegraphics[scale=0.75]{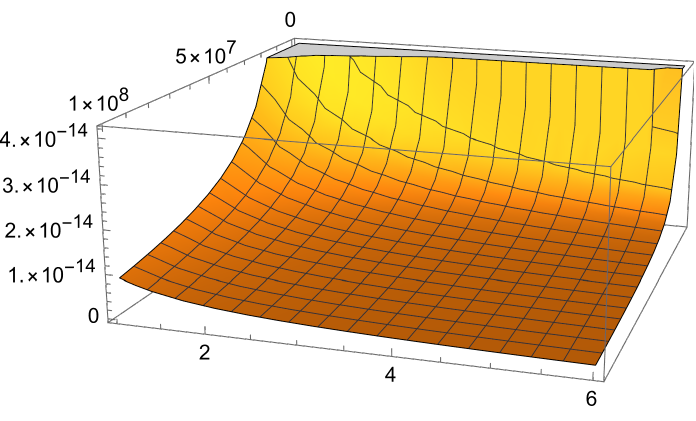}
\caption{Self-mirror scalar potential $V$ according to $\tau_0$ and $\tau_1$ with numerical constant value of 
$W_0=1, A_0=1,  a_0= 2 \pi, b_0= 1/4, \tau_1=10^7$. The potential ends with positive value approaching the large volume limit.}
\label{figure:SPotential-QS1}
  \end{figure}

Regarding $V=V(\mc{V}, \tau_0)$, the de Sitter vacuum up to order $\mathcal{O} (\frac{1}{\mc{V}^{3}})$ can be solved as we have in \cite{Sun:2024ral}, via
\be\label{qv}
\frac{\partial V}{\partial \mc{V}}
=\frac{\partial V}{\partial \tau_0}=0,
\ee
and we propose the de Sitter vacuum essentially arise from the $\mathcal{O} (\frac{1}{\mc{V}^{2}})$ uplift term. 
The location of the minimum may be solved around a saddle point solution at the large volume limit from $\frac{\partial V}{\partial \tau_0}=0$ with\footnote{Note that the minimum solution of $\mc{V}$ from \eqref{qv} can also alternatively be solved through $\frac{\partial V} {\partial \mc{V}} = 0$ with solution may be plotted in a similar shape of effective scalar potential.}
\be \label{tau}
 \mc{V} =-\frac{e^{a_0 \tau_0} W_0}{3 a_0^2 A_0 \cos(a_0 b_0) \sqrt{\tau_0} }.
 \ee
It is obvious that while $\cos(a_0 b_0)<0$ and $a_0 b_0 \to  \frac{\pi}{2} K,  \frac{3\pi}{2} K, K \in \mbb{Z}_+$, the total volume goes to large volume limit.  
As follows, the derivative with respect to $\mc{V}$ becomes
\be 
\frac{\partial V} {\partial \mc{V}} = 0
 \Rightarrow
e^{a_0 \tau_0}+ 2A_0(1+a_0 \tau_0(1+a_0\tau_0)) \cos(a_0 b_0)=0,
\ee
in which $a_0 \tau_0 \gg 1$ is required to ignore higher order instanton corrections. The value of $\tau_0$ can therefore be solved into
\be\label{qtau}
\tau_0=- \frac{2}{a_0} ProductLog\left[-\frac{1}{2} 
\sqrt{-\frac{W_0 }{2 A_0 \cos(a_0 b_0)}}\right]>0,
\ee
where $ProductLog[z]$ represents the principal solution of $w$ in $z=w e^w$, and $\tau_0>0$ with $a_0>0, A_0>0, b_0>0$. 
Analytically, by inserting the solution \eqref{qtau} and \eqref{qv} back into the effective scalar potential at large volume limit \eqref{VSchoen001} one can verify that this indeed gives a positive vacuum solution as de Sitter vacuum.
Numerically, the de Sitter vacuum  can be illustrated at the saddle point region in Figure \ref{figure:SPotential-QS2}.
 \begin{figure}[h]
 	\centering
\includegraphics[scale=0.75]{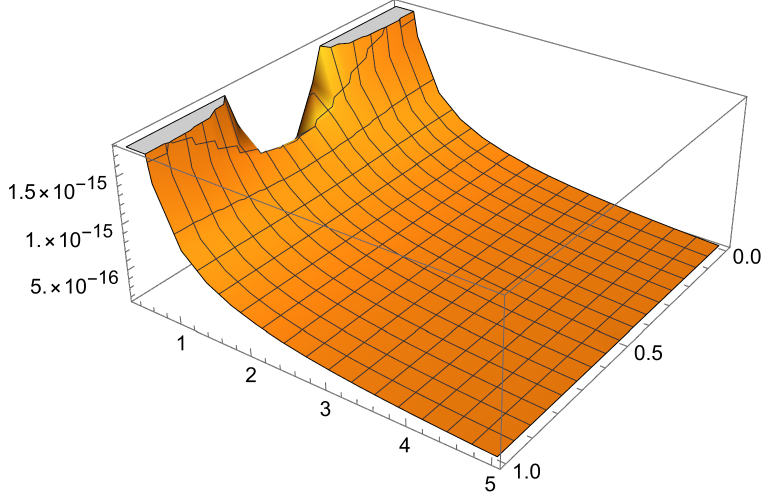}
\caption{Quotient Schoen scalar potential at the large volume limit $\tau_1, \tau_2\to \infty, a_0\tau_0\sim \ln \mc{V}$, according to $\tau_0$ and axionic field $b_0=[0, 1]$. Here the value $W_0=1,A_0=1,  a_0=2\pi, \tau_1= 10^8, \tau_2= 10^8$ are used.
The positive uplift term gives rise to de Sitter vacuum at saddle point region.}
\label{figure:SPotential-QS2}
\end{figure}

Although we have solved the minimum analytically up to the order  $\mathcal{O} (\frac{1}{\mc{V}^{3}})$, we propose that the de Sitter essentially given rise from the leading uplift term $\frac{W_0^2}{\mc{V}^2}$
which is in the same order as F-term $\frac{D W. DW}{\mc{V}^2}$ 
in $\mathcal{O} (\frac{1}{\mc{V}^{2}})$. 

\subsubsection{Large Volume Scenario II}

Consider the large volume limit $\mc{V} \to \infty$ with the base divisor $\tau_0$ to be exponentially large,
the non-perturbative terms contribute to the effective scalar potential such that
\be
W = W_0 + A_1 e^{-a_1 T_1} + A_2 e^{-a_2 T_2}.
\ee
The effective scalar potential \eqref{VQ} then results in 
\bea
V &=& \frac{W_0^2 \tau_0^6}{2916~\mc{V}^6} 
-\frac{\left(a_1 A_1 e^{-a_1 \tau_1} \cos(a_1 b_1) + a_2 A_2  e^{-a_2 \tau_2} \cos(a_2 b_2)\right)W_0 \tau_0^7}{104976 ~\mc{V}^{6}}  \\
\nonumber
& & -\frac{ W_0^2 (\tau_1+\tau_2)\tau_0^5}{2187~ \mc{V}^{6}}+\frac{\left(a_1^2 A_1^2 e^{-2 a_1 \tau_1}+ a_2^2 A_2^2 e^{-2 a_2 \tau_2}+ a_1 A_1 a_2 A_2 e^{-a_1 \tau_1-a_2 \tau_2}\cos(a_1 b_1- a_2 b_2)\right)\tau_0^8}{1889568~ \mc{V}^{6}},
\eea
with K\"ahler metric \eqref{Kmetric} read off from $i,j=1,2$. 
Consider that the total volume of quotient Schoen Calabi-Yau \eqref{vQSchoen} with $\tau_0\to \infty$ approaches 
\be
\mc{V}= {1\over 18}\sqrt{\tau_0 (\tau_0-6 \tau_1)(\tau_0-6 \tau_2)}
 ~\sim {1\over 18}~ \tau_0^{{3 \over 2}}, ~~
 \text{with}~ \tau_0 \sim (18\mc{V})^{2 \over 3}.
\ee
The effective scalar potential in the order of $\mathcal{O} (\frac{1}{\mc{V}^{n}})$ becomes
\bea \label{SV}  \nonumber
V &=& \frac{36 W_0^2}{\mc{V}^2} 
-\frac{ 8 (\frac{2}{3})^{1/3}W_0^2 (\tau_1+\tau_2)}{\mc{V}^{8/3}}
-\frac{2^{2/3} 3^{4/3}\left(a_1 A_1 e^{-a_1 \tau_1} \cos(a_1 b_1) + a_2 A_2  e^{-a_2 \tau_2} \cos(a_2 b_2)\right)W_0 }{\mc{V}^{4/3}}  \\
& & 
+2^{1/3} 3^{2/3}\frac{\left(a_1^2 A_1^2 e^{-2 a_1 \tau_1}+ a_2^2 A_2^2 e^{-2 a_2 \tau_2}+ a_1 A_1 a_2 A_2 e^{-a_1 \tau_1-a_2 \tau_2}\cos(a_1 b_1- a_2 b_2)\right)}{\mc{V}^{2/3}} .
\eea
Under the convenient decompactification limit with
\eq{
a_1 A_1=W_0,~~ a_2 A_2=W_0,
}
the structure of effective scalar potential \eqref{SV}, at large volume limit $\tau_0\to \infty$, gets clear in
\bea
V& =&\frac{36 W_0^2}{\mc{V}^2} -\frac{\kappa (\tau_1+\tau_2)}{\mc{V}^{8/3}}
- \frac{\lambda(e^{-a_1 \tau_1} \cos(a_1 b_1) + e^{-a_2 \tau_2} \cos(a_2 b_2))}{\mc{V}^{4/3}}
\\ \nonumber
& & +\frac{\mu (e^{-2 a_1 \tau_1}+ e^{-2 a_2 \tau_2}+ e^{- a_1 \tau_1- a_2 \tau_2}    \cos(a_1 b_1- a_2 b_2)}{\mc{V}^{2/3}},
\eea
with $\kappa= 8 (\frac{2}{3})^{1/3} W_0^2$, $\lambda=2^{2/3} 3^{4/3}W_0^2$, $\mu=2^{1/3} 3^{2/3}W_0^2$.
To suppress the higher order instanton correction, $a_1 \tau_1, a_2 \tau_2\gg 1$ shall be set. By choosing the decompactification limit with the base and fiber divisor following
\be
\tau_0 \to \infty, ~~\text{with}~~ a_1 \tau_1\sim\ln \mc{V},~ a_2 \tau_2\sim\ln \mc{V},
\ee
the scalar potential in the order of $\mathcal{O} (\frac{1}{\mc{V}^{n}})$ finally results in
\be
V =\frac{36 W_0^2}{\mc{V}^2} 
- \frac{\lambda( \cos(a_1 b_1) +  \cos(a_2 b_2))}{\mc{V}^{7/3}}
 +\frac{\mu (2+ \cos(a_1 b_1- a_2 b_2)}{\mc{V}^{8/3}}
 -\frac{\kappa (\tau_1+\tau_2)}{\mc{V}^{8/3}},
\ee
Here it is obvious that the first term $36 W_0^2 \over \mc{V}^2$ takes the dominant value which is obviously positive in the order of $\mathcal{O} (\frac{1}{\mc{V}^{2}})$, and the subleading terms suppressed at the order of $\mathcal{O} (\frac{1}{\mc{V}^{7/3}})$ and $\mathcal{O} (\frac{1}{\mc{V}^{8/3}})$. Therefore, in total, the non-perturbative scalar potential takes a \emph{positive} value while $W_0$ stabilized by the complex structures and axio-dilatons from flux compactification. 
The double elliptic self-mirror construction of quotient Schoen Calabi-Yau also contributes positively to the scalar potential in the same order as F-term $\frac{D W. DW}{\mc{V}^2}$ in $\mathcal{O} (\frac{1}{\mc{V}^{2}})$. Such that, positive de Sitter vacuum approaches in the large volume limit.

Regard the effective scalar potential \eqref{SV} as $V=V(\mc{V}, \tau_1, \tau_2)$, the location of the de Sitter minimum shall be solved at 
\be \label{minimum}
\frac{\partial V} {\partial \mc{V}}
= \frac{\partial V} {\partial \tau_1}
= \frac{\partial V} {\partial \tau_2}=0.
\ee
Thanks to the  symmetric construction of volume form according to the fiber divisors $\tau_1$ and $\tau_2$, one can solve \eqref{minimum} in a hyperbolic way for $\frac{\partial V} {\partial \tau_1}= \frac{\partial V} {\partial \tau_2}=0$ 
with 
\be\label{V23}
\mathcal{V}^{2/3}= \frac{2^{1/3} 3^{2/3} W_0 (a_1^2 A_1 x \cos(a_1 b_1)- a_2^2 A_2  y\cos(a_2 b_2))}
{2 a_1^3 A_1^2 x^2 -2 a_2^3 A_2^2 y^2+(a_1-a_2) a_1 A_1 a_2 A_2 x y \cos(a_1 b_1-a_2 b_2)},
\ee
where $x:=e^{-a_1\tau_1},~ y:=e^{-a_2\tau_2}$. 
The value of $\tau_1, \tau_2$ may be further solved with the relation of $y$ and $x$, 
\be
y=-\frac{\kappa +2 \mu a_1 x^2 \mc{V}^2- \lambda a_1  \cos(a_1 b_1) x\mc{V}^{4/3}}{\mu a_1 \cos(a_1b_1-a_2b_2)x \mc{V}^2}>0,
\ee 
with negative value of $\cos(a_1b_1-a_2b_2)$ according to the leading order of $\mc{V}$,
and $\mathcal{V}^{2/3}$ \eqref{V23} under the decompactification limit $a_1 A_1=W_0, a_2 A_2=W_0$,
\be 
\mathcal{V}^{2/3}=\frac{2^{1/3} 3^{2/3} (a_1 x \cos(a_1 b_1)- a_2  y \cos(a_2 b_2))}
{2 a_1 x^2 -2 a_2 y^2+(a_1-a_2) x y \cos(a_1 b_1-a_2 b_2)}.
\ee
Although the de Sitter minimum can be analytically solved up to the order $\mathcal{O} (\frac{1}{\mc{V}^{8/3}})$, we still propose that the de Sitter minimum essentially arises from the leading uplift term in the order $\mathcal{O} (\frac{1}{\mc{V}^{2}})$.
Numerically, the de Sitter solution at the saddle point can be illustrated in  Figure \ref{figure:SPotential-QS3} at the large volume limit as we have shown in \cite{Sun:2024ral}. 
In addition, different solution of stable de Sitter vacuum at large volume limit with $\tau_0\to 0$ is also illustrated in Figure \ref{figure:SPotential-QS4}.
Note that beyond the region plotted, along the large volume direction with $\tau_0\to \infty$, the effective potential $V\to 0$ from above while the total volume $\mc{V} \to \infty$. Namely, at the large volume limit, the effective scalar potential gives rise to the de Sitter vacuum with a small positive value of scalar potential.
 \begin{figure}[h]
 	\centering 	\includegraphics[scale=0.7]{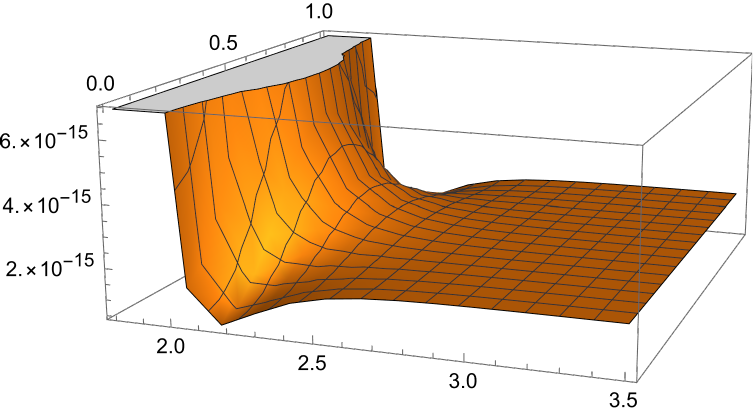}
	\caption{Quotient Schoen scalar potential at the large volume limit $\tau_0 \to \infty$ while $a_1\tau_1, a_2\tau_2 \sim \ln \mc{V}$. $V$ is plotted as function of $\tau_2$ and axionic field $b_2=[0, 1]$ according to \eqref{SV} with $ W_0=1, A_1=1, A_2=1, a_1= 2\pi, a_2=2 \pi, \tau_0= 10^6, \tau_1= 3, b_1=1$. 
  The de Sitter vacuum located along the $\tau_1, \tau_2$ directions as a saddle point.}
 \label{figure:SPotential-QS3}
  \end{figure}
 \begin{figure}[h]
 	\centering 	\includegraphics[scale=0.7]{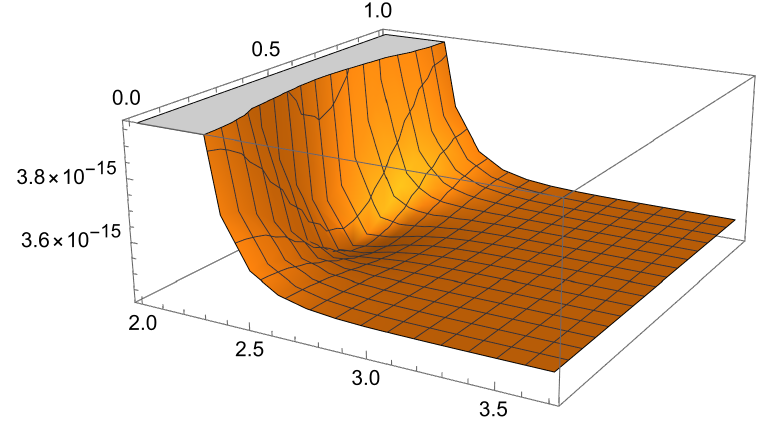}
	\caption{Quotient Schoen scalar potential at the large volume limit $\tau_0 \to \infty$ while $a_1\tau_1, a_2\tau_2 \sim \ln \mc{V}$. $V$ is plotted as function of $\tau_2$ and axionic field $b_2=[0, 1]$ according to \eqref{SV} with $ W_0=1, A_1=1, A_2=1, a_1= 2\pi, a_2=2 \pi, \tau_0= 10^6, \tau_1= 2, b_1=1$. 
  Different solution of stable de Sitter vacuum located along the $\tau_1, \tau_2$ directions.}
 \label{figure:SPotential-QS4}
  \end{figure}

\subsection{Schoen Calabi-Yau in Explicit}
\label{I}

Explicitly for Schoen Calabi-Yau threefold, $h^{1,1}=h^{2,1}=19$. 
In general, $19$ number of K\"ahler moduli lead to difficulties of studying the large volume scenario in the K\"ahler moduli space. 
However, thanks to the symmetric double elliptic construction,  the non-ambient two-cycle volumes can be related to the ambient ones via \eqref{eqnbasiswithHa}. This allows us to derive the total volume in the K\"ahler moduli space with the solution given in \eqref{vSchoen-s}. 

To obtain the non-perturbative terms, we first compute the inverse of K\"ahler metric as
\eq{\label{KSmetric}
G_n^{i\bar{j}}=
\begin{pmatrix}
(\tau_{n}^{(1)})^2+{\tau_{n}^{(1)} \tau_0}+\frac{\tau_0^2}{2} & {\tau_0^2 \over 4} & \qquad -{\tau_0^2 \over 2}\\
\newline\\
{\tau_0^2 \over 4} & (\tau_{n}^{(2)})^2+{(\tau_{n}^{(2)}) \tau_0}+\frac{\tau_0^2}{2} & \qquad -{\tau_0^2 \over 2}\\
\newline\\
-{\tau_0^2 \over 2} & -{\tau_0^2 \over 2} & \qquad \tau_0^2
\end{pmatrix},
} where $n=0,...8$ represents the summation  of $9$ fiber divisors into one from each of the double elliptic fibers. 
This sums up $9$ fiber divisors into one as two overall fiber K\"ahler moduli  $\tau_n^{(1)}$ and $\tau_n^{(2)}$, and allows us to preserve the no-scale structure with $G_n^{i\bar{j}} \partial_i {\cal K}~\partial_{\bar{j}} {\cal K}\equiv 3$.

\subsubsection{Large Volume Scenario I}

In the fist large volume limit, we set the fiber K\"ahler moduli $\tau_{n}^{(1)}, \tau_{n}^{(2)}$ both to be exponentially large and the base divisor $\tau_0$ with a finite value.
The non-perturbative contribution then reads
\be
W = W_0 + A_0 e^{-a_0 T_{0}}.
\ee
The involved inverse of K\"ahler metric element can be read off from \eqref{KSmetric} with $G^{00}=\tau_0^2$. And the self-mirror non-perturbative effective scalar potential \eqref{VQ} then results in
\be\label{VSchoen0}
V= \frac{W_0^2 (\tau_0 \tau_{n}^{(1)} \tau_{n}^{(2)})^2}{\mc{V}^6} 
+\frac{ 2 W_0^2 \tau_0^3 \tau_{n}^{(1)} \tau_{n}^{(2)}(\tau_{n}^{(1)}+\tau_{n}^{(2)})}{\mc{V}^{6}}
+\frac{2 A_0 W_0 e^{-a_0 \tau_0} (\tau_0 \tau_{n}^{(1)} \tau_{n}^{(2)})^2\cos(a_0 b_0)(1+a_0 \tau_0)}{\mc{V}^{6}}.
\ee
The total volume of Schoen Calabi-Yau \eqref{vSchoen-s} in the large volume limit behaves as
\bea
  {\cal V} &=& 
  {1\over 2}\sum_{i,j=0}^{8}\sqrt{\tau_0^3+2 \tau_0^2(\tau_{i}^{(1)}+\tau_{j}^{(2)})+ 4 \tau_0 \tau_{i}^{(1)} \tau_{j}^{(2)}} \\ \nonumber
 &\sim & \sum_{i,j=0}^{8}\sqrt{ \tau_0\tau_{i}^{(1)} \tau_{j}^{(2)}}, ~~
 \text{with}~ \tau_0 \tau_{n}^{(1)} \tau_{n}^{(2)} \sim \mc{V}^{2}.
\eea
Consider that K\"ahler moduli $\tau_n^{(1)}, \tau_n^{(2)}$ taking an equal position in the moduli space, the scalar potential \eqref{VSchoen0} in the order of $\mathcal{O} (\frac{1}{\mc{V}^{n}})$ reduces to 
\bea\label{Sv}
V &=& \frac{W_0^2}{\mc{V}^2} 
+\frac{ 4 W_0^2 \tau_0^{3/2}}{ \mc{V}^{3}}
+\frac{2 A_0 W_0 e^{-a_0 \tau_0} \cos(a_0 b_0)(1+a_0 \tau_0)}{\mc{V}^{2}}.
\eea
The subleading term with $\cos(a_0 b_0)$ is suppressed with axionic field $b_0$. While considering the large volume limit with $\mc{V} \to \infty,~ \tau_1, \tau_2 \gg \tau_0$, and
\be
a_0 \tau_0 \sim \ln \mc{V}, 
\ee
the effective scalar potential in the order of $\mathcal{O} (\frac{1}{\mc{V}^{n}})$ reduces to the structure
\bea\label{VSchoen01}
V &=& \frac{W_0^2}{\mc{V}^2} 
+\frac{ 4 W_0^2 \tau_0^{3/2}}{ \mc{V}^{3}}
+\frac{2 A_0 W_0  \cos(a_0 b_0)(1+\ln \mc{V})}{\mc{V}^{3}}.
\eea
Similar to the quotient Schoen large volume scenario, the Schoen large volume scenario also obtain a clear uplift term $\frac{W_0^2}{\mc{V}^2}$ in the order of F-term $\frac{D W. DW}{\mc{V}^2}$. 
The scalar potential approaches zero from above as the first term $\frac{W_0^2}{\mc{V}^2}$ dominates as illustrated in Figure \ref{figure:SPotential-S1} according to \eqref{VSchoen0}.
\begin{figure}[h]
 	\centering 	\includegraphics[scale=0.75]{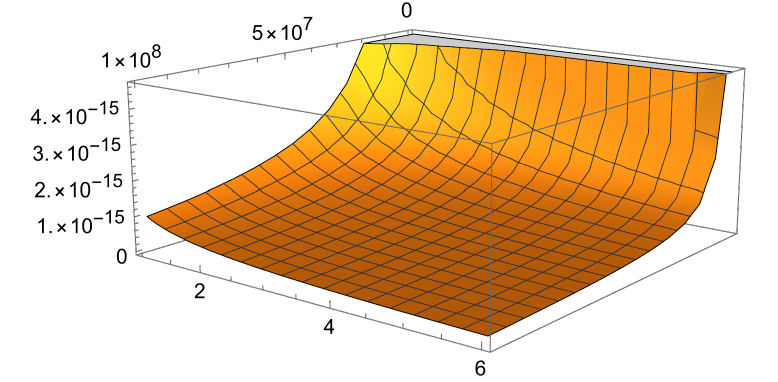}
\caption{Self-mirror scalar potential $V$ according to $\tau_0$ and $\tau_1$ with numerical constant value of 
$W_0=1, A_0=1,  a_0= 2 \pi, b_0= 1/4, \tau_n^{(1)}=10^7$. The potential ends with positive value approaching the large volume limit as well.}
\label{figure:SPotential-S1}
  \end{figure}
Regard $V=V(\mc{V}, \tau_0)$, the metastable de Sitter vacuum up to order $\mathcal{O} (\frac{1}{\mc{V}^{3}})$, can be solved at a saddle point with
\be
\frac{\partial V}{\partial \mc{V}}
=\frac{\partial V}{\partial \tau_0}=0.
\ee
The location of the minimum may be solved with a hyperbolic-like solution at the large volume limit from $\frac{\partial V}{\partial \mc{V}}=0$ with
\be \label{sv}
 \mc{V} =-\frac{6 e^{a_0 \tau_0} W_0 \tau_0^{3/2}}{  e^{a_0 \tau_0} W_0+ 2 A_0 \cos(a_0 b_0)(1 +  a_0 \tau_0) }.
 \ee
It is obvious that while $\cos(a_0 b_0)<0$ and $e^{a_0 \tau_0} W_0+ 2 A_0 \cos(a_0 b_0)(1 +  a_0 \tau_0)\to 0$, the total volume goes large volume limit.  
As follows, the derivative with respect to $\tau_0$ becomes
\be 
\frac{\partial V} {\partial \tau_0} = 0
 \Rightarrow
e^{a_0 \tau_0} W_0 + 2A_0(1+a_0 \tau_0(1+a_0\tau_0)) \cos(a_0 b_0)=0,
\ee
in which $a_0 \tau_0 \gg 1$ shall be required to suppress the higher order instanton corrections. The value of $\tau_0$ can again be solved as
\be\label{Stau}
\tau_0=- \frac{2}{a_0} ProductLog\left[-\frac{1}{2} 
\sqrt{-\frac{W_0 }{2 A_0 \cos(a_0 b_0)}}\right]>0,
\ee
and $\tau_0>0$ with $a_0>0, A_0>0, b_0>0$. 
Analytically, by inserting the solution \eqref{Stau} and \eqref{sv} back into the effective scalar potential at large volume limit \eqref{VSchoen01} one can verify that this indeed gives a positive vacuum solution as de Sitter vacuum.
Numerically, the metastable de Sitter vacuum  can be illustrated at the saddle point region in Figure \ref{figure:SPotential-S2}.
The  de Sitter vacuum essentially arises from the uplift term $\frac{W_0^2}{\mc{V}^2}$ which is
in the same order of F-term again.
 \begin{figure}[h]
 	\centering
\includegraphics[scale=0.7]{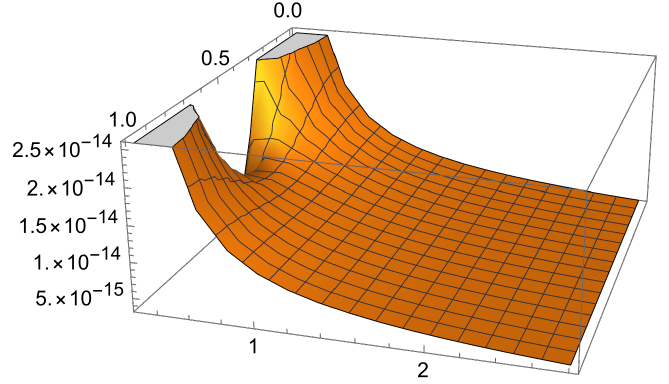}
\caption{Effective scalar potential from Schoen large volume scenario with
 $\tau_n^{(1)}, \tau_n^{(2)}\to \infty, a_0\tau_0\sim \ln \mc{V}$, according to $\tau_0$ and axionic field $b_0=[0, 1]$. Here value $W_0=1,A_0=1,  a_0=2\pi, \tau_1= 10^7, \tau_2= 10^7$ are used.
The positive de Sitter vacuum supposed to arises at the saddle point region.}
\label{figure:SPotential-S2}
\end{figure}

\subsubsection{Large Volume Scenario II}

In the alternative large volume limit, 
set $\tau_n^{(1)}$ and $\tau_n^{(2)}$  correspond to the sum of double elliptic fiber divisors into $D_1$ and  $D_2$ which can be embedded into a Calabi-Yau fourfold with elliptic torus fibration over a base threefold.
The non-perturbative contributions are manifested in the superpotential as 
\be
W = W_0 + A_1 e^{-a_1 T_{n}^{(1)}} + A_2 e^{-a_2 T_{n}^{(2)}},
\ee  
in which $T_{n}^{(1)}= \tau_n^{(1)}+ i b_n^{(1)},~ T_{n}^{(2)}= \tau_n^{(2)}+ i b_n^{(2)}$.
The non-perturbative effective scalar potential \eqref{VQ} with K\"ahler metric \eqref{KSmetric}, can be represented by
{\small 
\bea \label{VS2}
&V& = \frac{3W_0^2 \tau_0^6}{8~\mc{V}^6} 
+\frac{3\left(a_1 A_1 e^{-a_1 \tau_n^{(1)}} \cos(a_1 b_1) + a_2 A_2  e^{-a_2 \tau_n^{(2)}} \cos(a_2 b_2)\right)W_0 \tau_0^7}{16 ~\mc{V}^{6}}  \\
\nonumber
& &+ \frac{ W_0^2 (\tau_n^{(1)}+\tau_n^{(2)})\tau_0^5}{~ \mc{V}^{6}}+\frac{\left( a_1^2 A_1^2 e^{-2a_1 \tau_n^{(1)}}+  a_2^2 A_2^2 e^{-2a_2 \tau_n^{(2)}}+ a_1 A_1 a_2 A_2 e^{-a_1 \tau_n^{(1)}-a_2 \tau_n^{(2)}} \cos(a_1 b_1- a_2 b_2)\right)\tau_0^8}{32~ \mc{V}^{6}} .
\eea
}
In the large volume limit $\mc{V} \to \infty, \tau_0 \gg \tau_i^{(1)}, \tau_j^{(2)}$, the total volume of Schoen Calabi-Yau \eqref{vSchoen-s} behaves as
\bea
\mc{V} &=& {1\over 2}\sum_{i,j=0}^{8}\sqrt{\tau_0 (\tau_0+2 \tau_{i}^{(1)})(\tau_0+2 \tau_{j}^{(2)})}\\ \nonumber
 & &\text{with}~ 
 \sqrt{\tau_0 (\tau_0+2 \tau_{n}^{(1)})(\tau_0+2 \tau_{n}^{(2)})}=2\mc{V},
 ~\text{and}~
 { \tau_0 \sim (2\mc{V})^{2 \over 3}},
\eea
and thus the scalar potential reduces to 
{\small 
\bea \label{Vss} \nonumber
V &=& \frac{6 W_0^2}{\mc{V}^2} 
+\frac{8\cdot 2^{1/3}~  W_0^2 (\tau_{n}^{(1)}+\tau_{n}^{(2)})}{\mc{V}^{8/3}}
+   \frac{3\cdot 2^{2/3} \left(a_1 A_1 e^{-a_1 \tau_{n}^{(1)}} \cos(a_1 b_1) + a_2 A_2  e^{-a_2 \tau_{n}^{(2)}} \cos(a_2 b_2)\right)W_0 }{\mc{V}^{4/3}}\\
& &+
\frac{ 2^{1/3} \left(a_1^2 A_1^2 e^{-2a_1 \tau_{n}^{(1)}}+  a_2^2 A_2^2 e^{-2a_2 \tau_{n}^{(2)}}+ a_1 A_1 a_2 A_2 e^{-a_1 \tau_{n}^{(1)}-a_2 \tau_{n}^{(2)}} \cos(a_1 b_1- a_2 b_2)\right)}{\mc{V}^{2/3}} .
\eea
}
In the large volume limit $\mc{V} \to \infty,~\tau_0 \gg \tau_1, \tau_2$, by taking the decompactification limit  
\eq{
a_1 \tau_n^{(1)} \sim \ln \mc{V},~ a_2 \tau_n^{(2)} \sim \ln \mc{V},
}
the non-perturbative effective scalar potential further reduces to
\bea 
&V &= \frac{6 W_0^2}{\mc{V}^2} 
+ \frac{3\cdot 2^{2/3} \left(a_1 A_1  \cos(a_1 b_1) + a_2 A_2   \cos(a_2 b_2)\right)W_0 }{\mc{V}^{7/3}}\\
\nonumber
& &+
2^{1/3}~ \frac{ a_1^2 A_1^2+ a_2^2 A_2^2+ a_1 A_1 a_2 A_2 \cos(a_1 b_1- a_2 b_2)}{\mc{V}^{8/3}} 
+\frac{8\cdot 2^{1/3}~  W_0^2 (\tau_{n}^{(1)}+\tau_{n}^{(2)})}{\mc{V}^{8/3}} .
\eea
Similar to the quotient Schoen Calabi-Yau large volume scenario, the leading term in the effective potential  $W_0^2 \over \mc{V}^2$  is positive  and leads to de Sitter vacuum at the large volume limit, while $W_0$  stabilized by the complex structures and axio-dilatons. 
The effective potential according to \eqref{VS2} can be illustrated as a function of divisor  $\tau_0$ and one of the fiber divisor sum $\tau_n^{(2)}$ as shown in Figure  \ref{figure:SPotential-S3}.
\begin{figure}[h]
 	\centering 	\includegraphics[scale=0.75]{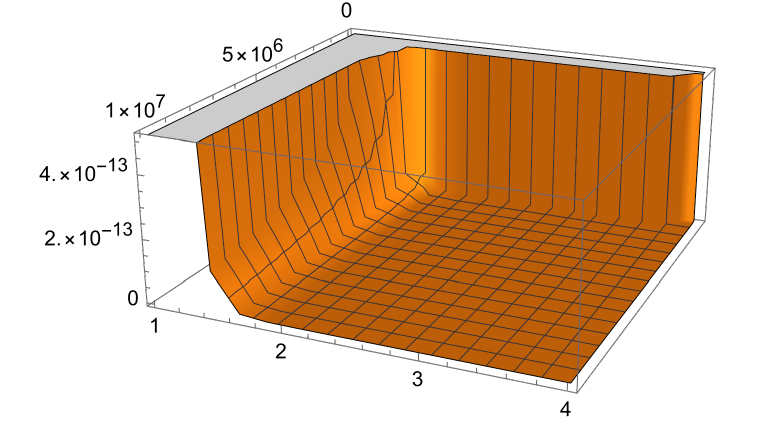}
\caption{Effective scalar potential according to $\tau_0\to \infty$ and $\tau_n^{(2)}$ with numerical constant value of 
$W_0=1, A_1=1, A_2=1, a_1= 2\pi, a_2=2 \pi, b_1= 1/2, b_2= 1/2, \tau_n^{(1)}=3$. The potential ends with positive value approaching the large volume limit.}
\label{figure:SPotential-S3}
  \end{figure}
The minimum of the effective scalar potential \eqref{Vss} $V=V(\mc{V}, \tau_{n}^{(1)}, \tau_{n}^{(2)})$ shall be solved at 
\be \label{minimum2}
\frac{\partial V} {\partial \mc{V}}
= \frac{\partial V} {\partial \tau_n^{(1)}}
= \frac{\partial V} {\partial \tau_n^{(2)}}=0.
\ee
Denoting $x:=e^{-a_1\tau_{n}^{(1)}},~ y:=e^{-a_2\tau_{n}^{(2)}}$, the volume of the Schoen Calabi-Yau threefold can be solved from $\frac{\partial V} {\partial \tau_1}= \frac{\partial V} {\partial \tau_2}=0$ and results in
\be\label{SV23}
\mathcal{V}^{2/3}= -\frac{3 \cdot 2^{1/3}  W_0 (a_1^2 A_1 x \cos(a_1 b_1)- a_2^2 A_2  y\cos(a_2 b_2))}
{2 a_1^3 A_1^2 x^2 -2 a_2^3 A_2^2 y^2+(a_1-a_2) a_1 A_1 a_2 A_2 x y \cos(a_1 b_1-a_2 b_2)},
\ee
with the symmetric construction of volume form according to the fiber divisors $\tau_n^{(1)}$ and $\tau_n^{(2)}$ considered. 
The value of $\tau_n^{(1)}$ and $\tau_n^{(2)}$ may be further solved via the relation of $y$ and $x$, 
\be \label{sy}
y=-\frac{8 W_0^2 -2 a_1^3 A_1^2 x^2 \mc{V}^2- 3 \cdot 2^{1/3} a_1^2 A_1 W_0 x \cos(a_1 b_1) \mc{V}^{4/3}}{ a_1^2 A_1 a_2 A_2\cos(a_1b_1-a_2b_2)x \mc{V}^2}>0,
\ee 
with positive value of $\cos(a_1b_1-a_2b_2)$ according to the leading order of $\mc{V}$,.
Similar to the quotient Schoen Calabi-Yau large volume scenario, although the de Sitter minimum can be analytically solved up to the order $\mathcal{O} (\frac{1}{\mc{V}^{8/3}})$, the stable de Sitter minimum essentially arises from the leading uplift term  $\frac{6 W_0^2}{\mc{V}^{2}}$.
Numerically, the stable de Sitter solution along the $\tau_{n}^{(1)}, \tau_{n}^{(2)}$ directions can be illustrated in  Figure \ref{figure:SPotential-S4} at the large volume limit. 
 \begin{figure}[h!]
 \centering 	\includegraphics[scale=0.55]{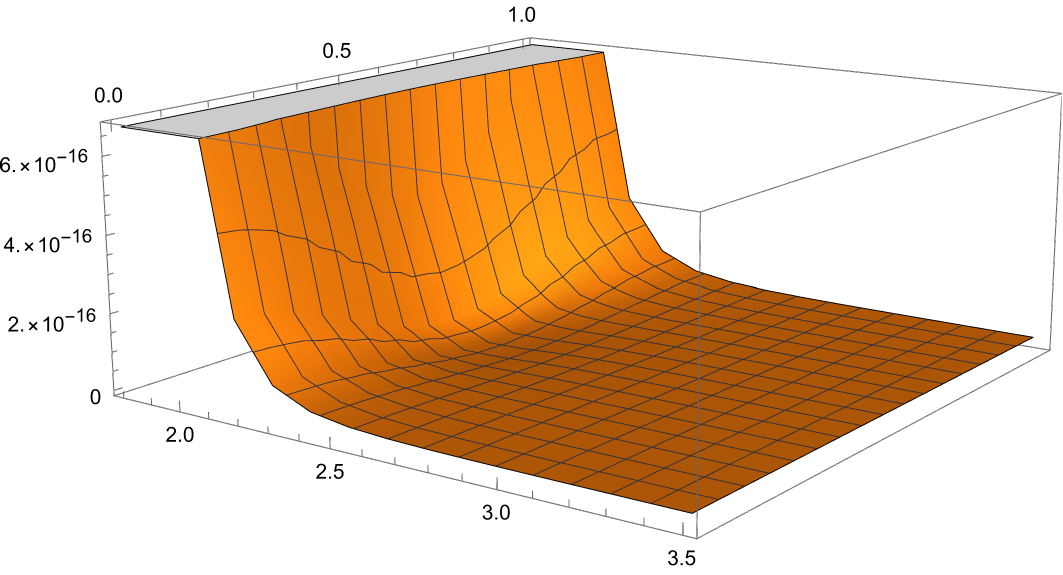}
	\caption{Effective scalar potential $V$  plotted as function of $\tau_{n}^{(2)}$ and axionic field $b_2=[0, 1]$ at large volume limit $\tau_0 \to \infty$ according to \eqref{VS2}. Here the value $W_0=1, A_1=1, A_2=1, a_1= 2\pi, a_2=2 \pi, \tau_{n}^{(1)}= 3, b_1=1/2, \tau_0= 10^6$ are used, while  stable de Sitter vacuum located at  $a_1\tau_{n}^{(1)}, a_2\tau_{n}^{(2)} \sim \ln \mc{V}$.}
 \label{figure:SPotential-S4}
  \end{figure}

\section{Phenomenology, Cosmology and Swampland Conjecture}

Take the self-mirror Schoen large volume scenario II as example, numerically with the value $W_0=1, A_1=1, A_2=1, a_1= 2\pi, a_2=2 \pi, \tau_{n}^{(1)}= 3, b_1=1/2, \tau_0= 10^6$, we have the  stable de Sitter vacuum located numerically around  $\tau_{n}^{(1)} \sim  3, \tau_{n}^{(2)} \sim  3.2$ following $a_1\tau_{n}^{(1)}, a_2\tau_{n}^{(2)} \sim \ln \mc{V}$ with $a_1=a_2=2 \pi$. Numerically inserting into the solution of total volume \eqref{SV23} and vacuum position \eqref{sv}, we have the total volume 
$\mc{V}=5.18\times 10^{11}$ in string units, with the de Sitter vacuum located at $a_1\tau_{n}^{(1)}, a_2\tau_{n}^{(2)} \sim  \mathcal{O}(10) \sim \ln \mc{V}$.
The values of the K\"ahler moduli found above together with the flux stabilized complex structure and axio-dilation of $W_0$ give rise to the minimum of the full scalar potential \eqref{Vscalar}. 
As follows, with the 4d Planck mass $M_p=1.22 \times 10^{19} \rm{GeV}$, the gravitino mass $m_{\frac{3}{3}}=e^{\mc{K}/2} W$ shall be given by $10^{-11} M_p=10^8 \rm{GeV}$ and the string scale results in $M_s=(M_p g_s)/\sqrt{\mc{V}}\sim 3 \times 10^{12} \rm{GeV}$ for $g_s=0.1$.
Such that an explicit realization of intermediate scale string scenario may be realized therein. 

On the cosmology aspect, we would like to emphasize that in our moduli stablization to de sitter vacua, we have included the axionic field $b_i$ dependences explicitly. 
These involved terms appear in cosine factors in precise. In particular, these terms involving in the realization of de Sitter vacua, also have implications for axion-like dark energy and dark matter candidates, such as weakly interacting slim particles for dark matter. 
It would be interesting to further study the axion potential as cosine functions together with the derivation of de Sitter vacua.

For swampland conjecture criteria, let's take the quotient Schoen large volume scenario II as an example. 
In the large volume limit with the moduli field $\tau_0$ to be exponentially large and $\tau_0 \gg \tau_1, \tau_2$, the 
total volume can be asymptotically represented via $\tau_0$ from $T_0\equiv \tau_0 +i b_0$, and the K\"ahler potential  $\mc{K} \sim -3 \ln \, (T_0+\bar{T}_0 )$.  
The kinetic energy from large moduli field $\tau_0$ is given by 
\be\label{kinetic}
 \frac{\partial^2 \mc{K}}{ \partial T_0\partial \bar{T}_0 }  ( \partial T_0  \partial \bar{T}_0)  = \frac{3}{4} \frac{(\partial \tau_0)^2}{\tau_0^2},
\ee
while the axionic field $b_0$ from the imaginary term is taken to be constant cosine factors, and therefore not enter into the derivative.
As the moduli fields shall have canonical normalized kinetic energy according to the swampland conjecture, one can rewrite the kinetic energy in terms of a scalar field $\Phi$ with canonical kinetic energy term for which $\Phi \equiv q \ln \tau_0$. The constant coefficient $q$ shall be determined such that $\Phi$ being canonical. Consider $\Phi$ provides  canonical kinetic energy term, it leads to 
\be
\frac{1}{2} (\partial \Phi)^2 = \frac{q^2}{2}  \frac{(\partial \tau_0)^2}{\tau_0^2},
\ee
which shall be equal to the right hand side of \eqref{kinetic}. Therefore, for the canonical redefinition 
\be
\frac{q^2}{2}= \frac{3}{4} ~~\Rightarrow~~q=\sqrt{3/2}\, ,
\ee shall be made. And such that
\be
\Phi= \sqrt{3/2} \ln \, \tau_0 \ .
\ee
Recall that the effective scalar potential of quotient Schoen large volume scenario II is lead by the dominant leading term $\frac{36 W_0^2}{\mc{V}^2}$, 
the effective scalar potential behaves as
\be
 V \sim \frac{36 W_0^2}{\mc{V}^2} 
 \propto\frac{1}{\tau_0^{3}}
\ee and therefore we can write the effective scalar potential as
\be
 V \propto e^ {-3 \sqrt{2/3} \Phi}.
\ee
As follows, the swampland conjecture criteria restrict as 
\be
\frac{|\nabla V|}{V} =\frac{|d V/d \Phi|}{V}= \sqrt{6} > \frac{2}{\sqrt{d-2}},
\ee
while $d=4$ is the spacetime dimension.
Therefore, in the large volume limit with the large field $\tau_0 \gg \tau_1, \tau_2$, the swampland lower bound of $\sqrt{2}$ in the large field limit is satisfied according to the transplanckian censorship conjecture  \cite{Bedroya:2019snp}.

Regarding to related swampland conjecture discussion, quotient Schoen Calabi-Yau manifolds are also studied in \cite{Deffayet:2023bpo, Deffayet:2024hug} from heterotic M-theory setting through F-term potential. It was also verified that potentials with stable vacua of positive, zero and negative vacuum energy from quotient Schoen compactification also fulfilled the swampland conjecture at large moduli field regime and at small value of moduli field near the center of the moduli space.

\section{Conclusion and Outlook}

Motivated by flux compactification with duality manifested, we studied the large volume scenario from self-mirror Calabi-Yau compactification.
In particular, we explicitly studied the 
geometry of self-mirror Schoen Calabi-Yau threefold with triple intersection given and total volume explicitly derived according to the K\"ahler moduli. 
The moduli space not only contains the ambient divisors but the non-ambient divisors as well. Namely, the two-cycle volume moduli corresponding to the K\"ahler moduli are not only the ones can be derived from hypersurface of higher dimension with toric methods, but the non-ambient two-cycle volumes incorporated as well. This is essentially managed with the relation of ambient and non-ambient classes from the double elliptic construction of Schoen manifold. 
Based on these, we then precisely presented particular approaches to reach the large volume limits with either the base or fiber divisor to be exponentially large.

By properly approaching the large volume limits of Schoen Calabi-Yau manifold, we then studied the self-mirror large volume scenario with effective scalar potentials derived.
Consider that the $\alpha'$ correction is naturally trivialized in the self-mirror large volume scenarios with trivial self-mirror Euler characteristic, we derived the effective scalar potential from the non-perturbative terms. 
In particular, at the large volume limit, the non-perturbative terms contribute to the scalar potential in the order of $\mc{O}\left(\frac{1}{\mc{V}^2}\right)$ in  K\"ahler moduli space, while the $\alpha'$-corrections are trivialized due to self-mirror Calabi-Yau construction. In total, at the large volume limit, the self-mirror Calabi-Yau compactification of Schoen type provides an order of $\mc{O}\left(\frac{1}{\mc{V}^2}\right)$ uplift term to the effective scalar potential with de Sitter vacuum.
Intriguingly, these de Sitter vacua derived from the self-mirror large volume scenario are essentially given by the leading positive uplift terms via dominant non-perturbative contribution. 
Moreover, the uplift term is in the same order as  F-term $\frac{D W. DW}{\mc{V}^2}$,
also same as the order of $W_0$ which is stabilized by the complex structure and axio-dilaton. 
We expect that the special dominant uplift term arises from the symmetric construction of the self-mirror Schoen Calabi-Yau manifolds, and therefore we propose a new mechanism for de Sitter uplift from self-mirror large volume scenario.

Moreover, note that although we obtained de Sitter vacua from self-mirror large volume scenarios, we are not proposing to violate the swampland conjecture but propose to introduce duality and mirror symmetry, such as through self-mirror calabi-Yau compactification, to derive de Sitter vacua. 
The natural embedding of T-duality in self-mirror Calabi-Yau manifolds allows geometric and non-geometric T-dual fluxes in one specific Calabi-Yau compactification and this may reveal new approaches of allowing de sitter vacuum with the special uplift term/mechanism.

For such self-mirror large volume scenario, it would be interesting to further investigate related cosmology model building, for example on the perspectives of early universe, dark matter and inflation models. 
Moreover, it is also interesting to study string compactification with mirror symmetry and T-duality (\emph{e.g.,} with geometric and non-geometric T-dual fluxes) for particle physics, realization of Standard Model and beyond.  
We hope to get back to these topics in the near future as well.

\acknowledgments

We would in particular like to thank Ralph Blumenhagen, Fernando Quevedo, Jie Zhou for many helpful comments, and thank
Michele Cicoli, Wolfgang Lerche, Andre Lukas, Dieter L\"ust,  Pramod Shukla for helpful discussions.
We would also like to thank Xiaoyong Chu, Chuying Wang, Xin Wang, Yaoxiong Wen, Lina Wu, Shing Tung Yau and Piljin Yi for helpful discussions on related topics.
RS is supported by KIAS New Generation Research Grant PG080701, PG080704, and acknowledge Ludwig Maximilian University of Munich, Max Planck Institute of Physics for their hospitality, and LMU-China Academic Network for their support when this research was initialized.


\bibliographystyle{JHEP}
\bibliography{reference}

\end{document}